\documentstyle[epsfig,amsmath,12pt]{article}

\newcommand{\ssup}{\raisebox{-.7ex}{$\stackrel{>}{\sim}$}}

\begin{document} 
\title{Systematics of spin-polarized small Na clusters}
\author{K. Andrae, P.-G. Reinhard, E. Suraud
 \\
        Institut f{\"u}r Theoretische Physik,\\
        Universit{\"a}t Erlangen,\\
        Staudtstrasse 7, D-91058 Erlangen, Germany\\
       and\\
         Laboratoire de Physique Th\'eorique, \\
        Universit{\'e} Paul Sabatier, \\
        118 Route de Narbonne, F-31062 Toulouse, cedex, \\
        France 
} 
\date{\today} 
\maketitle 
 
\abstract{ 
Inspired by recent experiments on fully spin polarized Na clusters, 
we perform a systematic survey of neutral Na clusters at all 
conceivable spin polarizations. We study the impact of spin state 
on ionic configuration, on global shape, and on optical response. 
For small clusters, the magic electronic shell at 4 spin-up electrons 
is a dominating feature leading to preferred binding for all clusters 
having four spin-up electrons (combined with 1 to 4 spin-down electrons). 
Such a preference fades away for larger systems where the 
unpolarized state is generally preferred. 
} 
 
 


\section{Introduction} 
 
Clusters in contact with a medium display a rich variety of 
possibilities not accessible with merely free clusters. A typical 
example is here a medium consisting out of a large drop of liquid 
$He$.  It provides a low-temperature laboratory for studies of various 
dynamical effects on small molecules and clusters \cite{toennies}. It 
has also been used as a tool to produce free Mg clusters 
\cite{doeppner}.  Accordingly, there exist several publications 
investigating the cluster-medium interaction, see 
e.g. \cite{leiderer,anciletto}.  A particularly interesting effect 
emerges, if Na is brought into contact with the drops. Sodium atoms 
can then form clusters on the surface and the high volatility of the 
drop acts as a criterion to select fully spin-polarized Na clusters 
\cite{Sti01}. A series of such clusters has been observed up to $N= 
16$ \cite{stienke}. We take these recent experiments as motivation for 
a theoretical survey of spin-polarized Na clusters.  The aim of this 
paper is to investigate from a theoretical perspective the structure 
and optical response of small neutral $Na $ clusters (up to $N = 10$) 
at all conceivable spin states. Selected examples of clusters with 
spontaneous spin-polarization had been studied in earlier publications 
\cite{magnresp,magnopt}. Here we aim at a more systematic survey going 
up to the extremes of fully spin-polarized systems.  We use the same 
tools as in previous publications, namely (time-dependent) 
density-functional theory for electrons and simulated annealing for 
ionic structure optimization.  The sequence of definite spins is 
produced by fixing the spin state of the cluster. The stability of 
fully spin-polarized configurations is checked by allowing the spins 
to vary freely.

\section{Formal framework} 
 
\subsection{Approach and computational scheme} 
 
The electron cloud of the clusters is described by density-functional
theory at the level of the local spin-density approximation (LSDA)
using the density functional of Ref. \cite{perwan} which is known to
perform reliably well also in the spin channel.  We complement that by
a self-interaction correction with the average-density approach
(ADSIC) \cite{adsic}. This provides correct ionization potentials and
an appropriate asymptotic tail of the density. The coupling of ions to
electrons is described by a local pseudo-potential which has been
proven to yield correct ground state and optical excitation properties
of simple sodium clusters \cite{kuemmel}. The spin-orbit force is
negligible for Na. This means that spin is totally decoupled from
spatial properties and there is rotational invariance in spin space.
It is to be expected that stationary states have collinear spin
distributions and integer total spin. We have checked that by allowing
non-collinear spin (for details see section \ref{sec:spinstab}) and we
find indeed collinear configurations throughout. Thus we continue with
computing the states with fixed total $z$-component of spin in
collinear configurations. In that cases, there is one spin orientation
throughout and its direction is arbitrary.  Clusters are characterized
then by their net polarization which is obtained as the sum of the
single z-component $S_z=s_z$. A coupling to good total spin produces a
sum of Slater states. This is what is called a correlated state. It
goes outside the realm of density functional theory.  Within LSDA we
have at hand only the total $z$-component $S=S_z$ and we take that to
characterize the spin state of the system. It is to be noted that it
is just that $S_z$ which couples to an external magnetic field. 
We thus have an appropriate measure of the magnetic response of the
cluster within LSDA, similar as it was studied before, e.g., in
\cite{magnresp}.

The electron 
wavefunctions are represented on an equidistant grid in 
three-dimensional coordinate space. The electronic ground state 
solution is found by iterative relaxation \cite{lauritsch}.  The 
electronic net spin is chosen at the initialization stage and stays 
unchanged throughout the relaxation. The ionic configurations are 
optimized by simulated annealing \cite{Metro}.  To compute the optical 
response, we propagate the electronic dynamics at fixed ionic 
configuration.  Propagation is done with time-dependent LSDA (TDLSDA) 
using the time-splitting method \cite{Fei82}. 
To compute the spectral 
distributions, we perform a spectral analysis after the TDLDA propagation 
\cite{yabber,bigtdlda}. This means that we initialize the dynamics by 
a small instantaneous boost of the center-of-mass of the electron 
cloud. The dipole moment is recorded during time evolution. A 
Fourier analysis into frequency domain then finally yields the spectral 
strength. For all technical details see the review \cite{ownrw}.

The global shape of the cluster is characterized by the r.m.s. radius 
and the dimensionless quadrupole moments $\alpha_{2\mu}$ recoupled to 
the total deformation $\beta$ and triaxiality $\gamma$. The various quantities read,  
\begin{equation}
\label{eq:shape} 
  r 
  = 
  \sqrt{\frac{\langle r^2\rangle}{N}} 
  \;,\; 
  \beta 
  = 
  \sqrt{\alpha_{20}^2+2\alpha_{22}^2} 
  \;,\; 
  \gamma 
  = 
  \mbox{atan}\frac{\sqrt{2}\alpha_{22}}{\alpha_{20}} 
  \;,\; 
   \alpha_{2m} 
  = 
  \frac{4\pi}{5}\frac{\langle r^2Y_{2m}\rangle}{Nr^2} 
  \;.
\end{equation} 
These equations can be read in two ways: for electrons,  
$\langle ...\rangle$ is  
the moment of the density $\rho_{\rm el}({\bf r})$ 
and $N\equiv N_{\rm el}$, while for ions, 
one considers the classical moment  
$\langle f({\bf r})\rangle=\sum_I f({\bf R}_I)$ 
and identifies $N\equiv N_{\rm ion}$. 
The leading and  
most robust quantity  is 
the r.m.s. radius. Its relative variation is limited as we will see. 
The total deformation $\beta$ is still a robust quantity showing, 
however, large variations in the range 0--1/3. The triaxiality 
$\gamma$ is more special. One has to keep in mind that it is 
well defined only for sufficiently large deformation, typically 
$\beta\ssup  1/10$. It is undefined for $\beta=0$ and 
only vaguely defined for small $\beta$. 
 
Variation of the spin composition is the major objective of that 
paper. We need a compact notation to characterize that and we will 
denote it by an upper index. The notation Na$_N^{N_\uparrow 
N_\downarrow}$ stands for a Na$_N$ cluster with $N_\uparrow$ spin-up 
electrons and $N_\downarrow$ spin-down electrons. Without loss of 
generality, we sort $N_\uparrow\geq N_\downarrow$. The total electron 
number combines to $N=N_\uparrow+N_\downarrow$. The spin state is 
characterized by the net spin $S=N_\uparrow-N_\downarrow$.  Note that 
we will count spin in units of $\hbar/2$ throughout this paper.

\subsection{Collinearity} 
\label{sec:spinstab}

The explicit treatment of spins allows still two options: one can 
assume that all spins are aligned (collinear) or one allows for 
non-collinearity as it is necessary, e.g.  for magnetic materials 
\cite{Koh99b}. Materials with negligible spin-orbit coupling, as 
e.g. Na, should have always collinear spin. Nonetheless, we have 
checked that in detail using the code of \cite{Koh99b}. For a handful 
of test cases, we initialize the electronic configuration in a 
non-collinear state and watch the evolution of that during 
ground-state iteration. We always find a quick convergence 
towards collinear configurations.  
\begin{figure} 
\centerline{\epsfig{figure=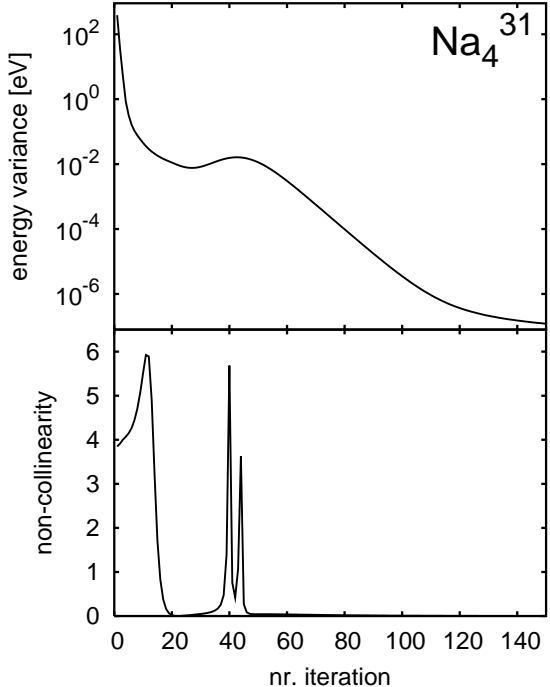,width=7.5cm}} 
\caption{\label{fig:noncol} 
Evolution of energy variance  
$\Delta E=  
\sqrt{\sum_\alpha(\varphi_\alpha|\hat{h}^2-\bar{h}^2|\varphi_\alpha)}$ 
(upper panel) and non-collinearity  
(lower panel) 
during iteration towards the electronic ground state configuration. 
The non-collinearity is defined in eq. (\ref{eq:noncol}). 
Test case of the cluster Na$_{4}^{(31)}$ in the final configuration 
with three spin-up and one spin-down electron. 
} 
\end{figure} 
One example is shown in figure \ref{fig:noncol}. The energy variance 
in the upper panel serves to demonstrate the overall convergence 
pattern. It remains to quantify the non-collinearity in one simple 
number. To that and, we evaluate the spin-orientation 
${\bf\sigma}_\alpha$ for each electron state $\alpha$ seperately, 
compute the angle between all pairs of electrons, and finally add up 
the quadratic deviations. This yields the ``non-collinearity'' as 
%
\begin{equation}
\label{eq:noncol} 
  \Delta\sigma^2 
  = 
  \sum_{\alpha\beta}\sin^2({\bf\sigma}_\alpha,{\bf\sigma}_\beta) 
  \;,\; 
  \sin^2({\bf a},{\bf b}) 
  = 
  1-\frac{({\bf a}\!\cdot\!{\bf b})^2}{{\bf a}^2{\bf b}^2} 
  \;,\;
  {\bf\sigma}_\alpha 
  = 
  (\varphi_\alpha|\hat{\bf\sigma}|\varphi_\alpha) 
  \;,
\end{equation} 
where $\hat{\bf \sigma}$ is the vector of Pauli spin matrices, 
i.e. the spin in units of $\hbar/2$. 
That measure is shown in the lower panel of figure \ref{fig:noncol}. 
One sees a quick convergence to collinearity, an interesting 
interludium where non-collinearity pops up again to foster a quick 
transition into a better configuration, and finally a stable 
collinear state. All cases studied showed the same stable final 
convergence to a collinear configuration. Thus we use the code 
with (much faster) thoroughly collinear handling of spin.

However, the non-collinear confgurations become important for 
detailed studies of spin stability. Each spin state is conserved for 
the present energy functional. It is only by perturbations, namely a 
small sin-orbit coupling in the pseudo-potentials or external magnetic 
fields, that the spin states can mix and undergo transitions which 
possibly run through non-collinear configuration as transitional 
stages. Such studies go beyond the scope and limitations of the 
present paper. We will adopt, in accordance with experimental claims 
\cite{stienke}, that the spin states once prepared stay stable for the 
necessary analyzing time.

\section{Results and Discussion} 
 
\subsection{Details of configurations and shapes} 
\label{sec:config} 
 
\begin{figure} 
\begin{center} 
\begin{picture}(164,290)(0,0) 
\put(43,228){\epsfig{figure=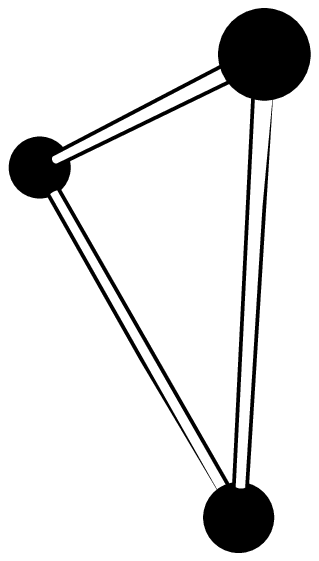,width=32pt}} 
\put(110,228){\epsfig{figure=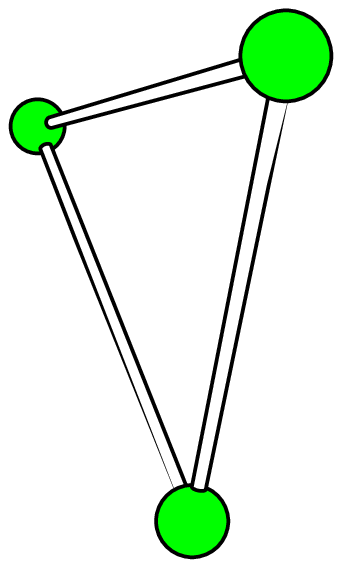,width=40pt}} 
\put(0,183){\epsfig{figure=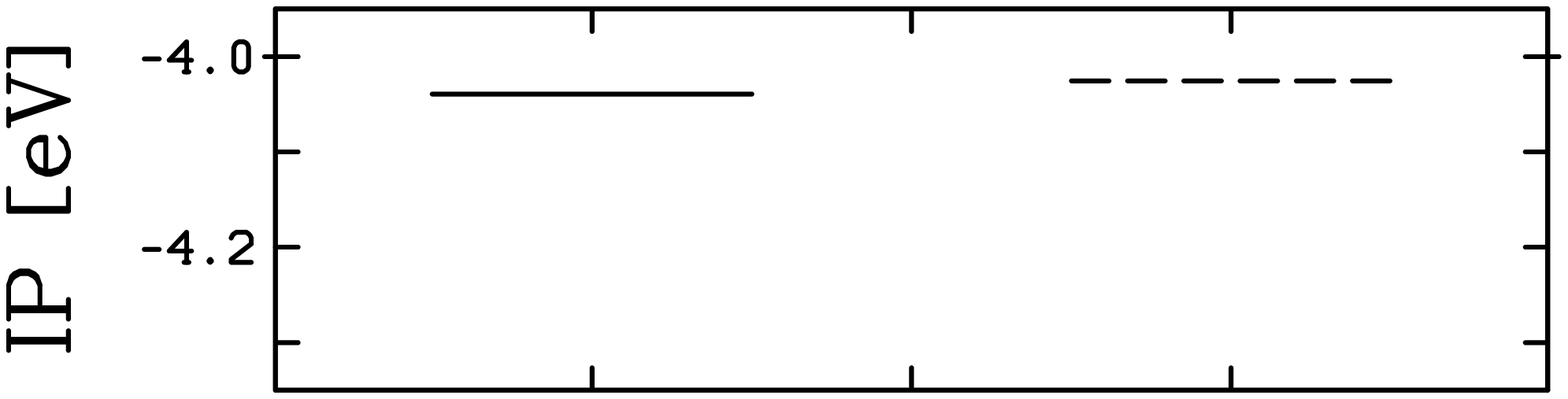,width=8.2cm}} 
\put(0,0){\epsfig{figure=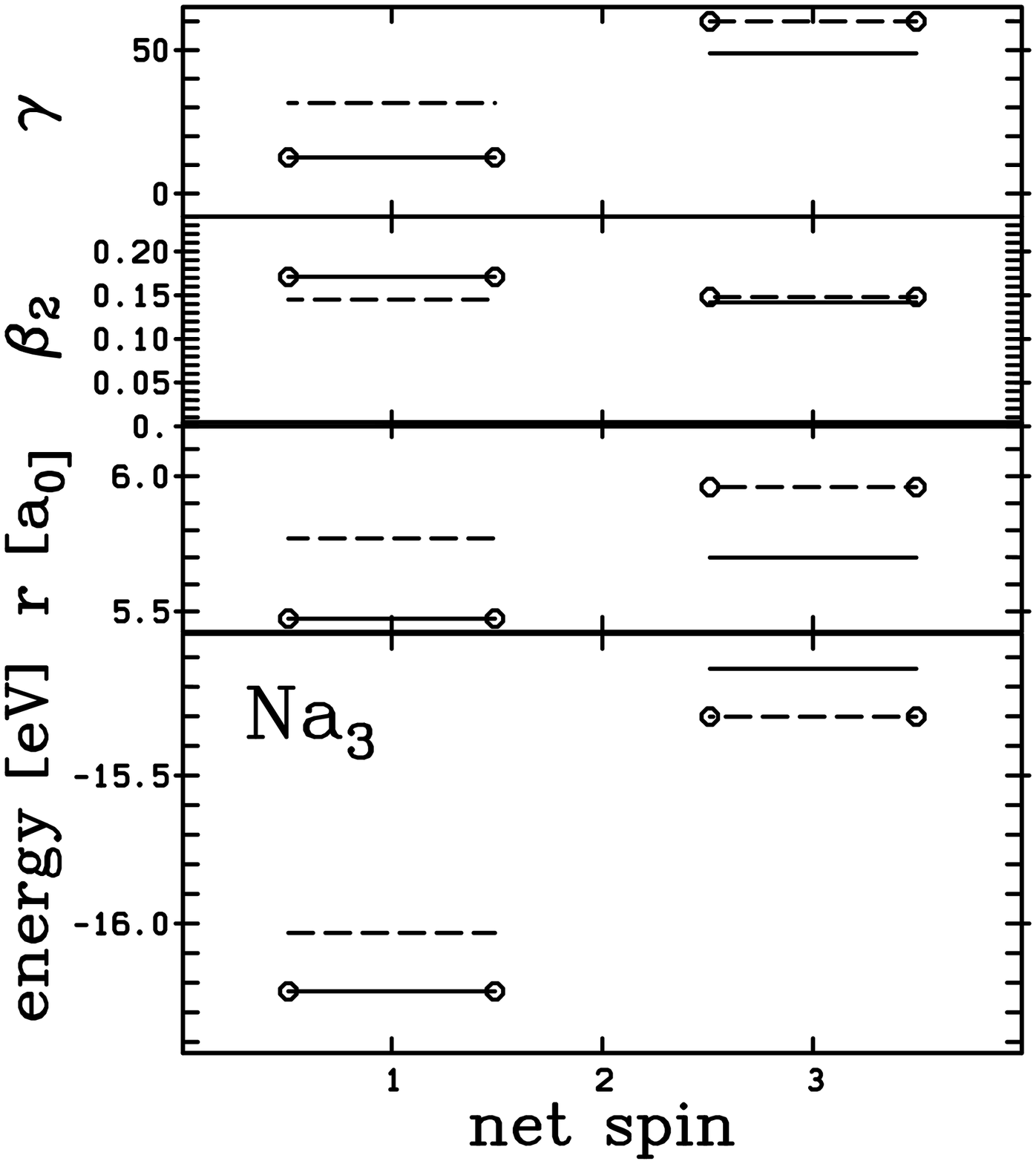,width=8.2cm}} 
\end{picture} 
\end{center} 
\caption{\label{fig:na3_detail}  
Global properties of Na$_3$ in various 
spin states: Lower panel = binding energy, second lower panel = 
r.m.s. radius, middle panel = total quadrupole deformation, 
second upper panel = triaxiality, uppermost panel = 
ionization potential (IP). The shape parameters are related to the 
electronic distribution and defined in eq. (\ref{eq:shape}).  On top 
of the panels, the ionic configurations which had been optimized for 
given electronic net spin are 
shown. The panels show results where  
all possible net-spins are computed in connection with all 
possible configurations.  The line type indicates the ionic 
configuration: full line $\leftrightarrow$  optimized for spin=1,  
dashed $\leftrightarrow$ optimized for spin=3.  
The (preferable) cases where 
actual electronic net-spin coincides with the spin for which the 
ionic configuration has been optimized are indicated by circles. 
The corresponding ionic configurations are shown on top of the panels. 
} 
\end{figure}

In a first round we investigate the structure of polarized Na clusters. For a 
given size $N$, we can have $N/2+1$ different spin states $S=N, N-2, N-4, ..., 
$ mod ($N,2$). For each spin state $S=N_\uparrow-N_\downarrow$ kept 
fixed, we optimize the ionic 
structure. As a consequence, we generate all ground state configurations for 
systems ($N,S$) with $N=3,..,8,10 $ and all $S$ as given above. One can now 
imagine that a given ground state for ($N,S$) may undergo a sudden spin flip 
$S \rightarrow S'$. The ions readjust to the new $S'$ very slowly. Electronic 
relaxation is much faster. We thus obtain a transient cluster with ionic 
structure of ($N,S$) but electronic structure readjusted to $S'$. The 
following figures show results for all the ground states as well as for all 
possible combinations to transient states $S \rightarrow S'$. The ionic 
configurations used are distinguished by line types, see figure captions. The 
actual electronic spin is given on the abscissa (in electronic spin unit 
($\hbar/2$)).

\begin{figure} 
\begin{center} 
\begin{picture}(164,290)(0,0) 
\put(33,228){\epsfig{figure=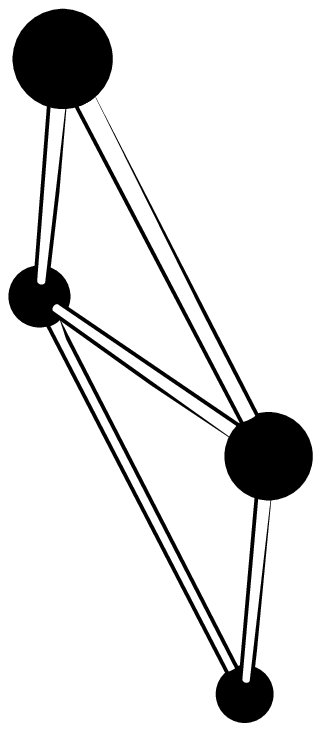,width=26pt}} 
\put(75,228){\epsfig{figure=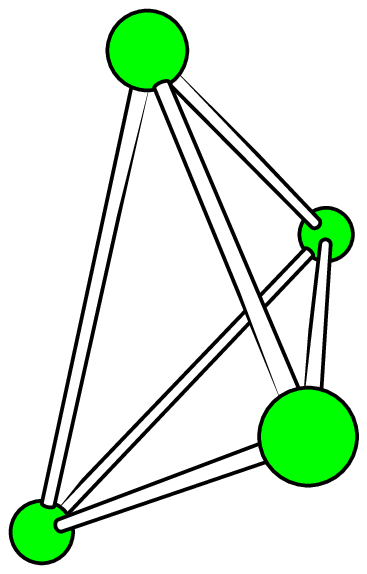,width=43pt}} 
\put(120,228){\epsfig{figure=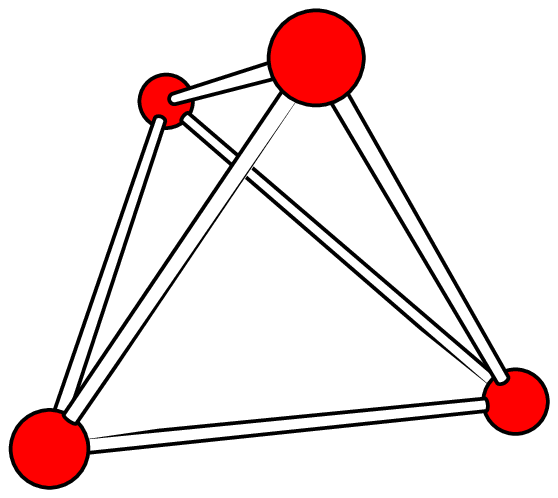,width=45pt}} 
\put(0,182){\epsfig{figure=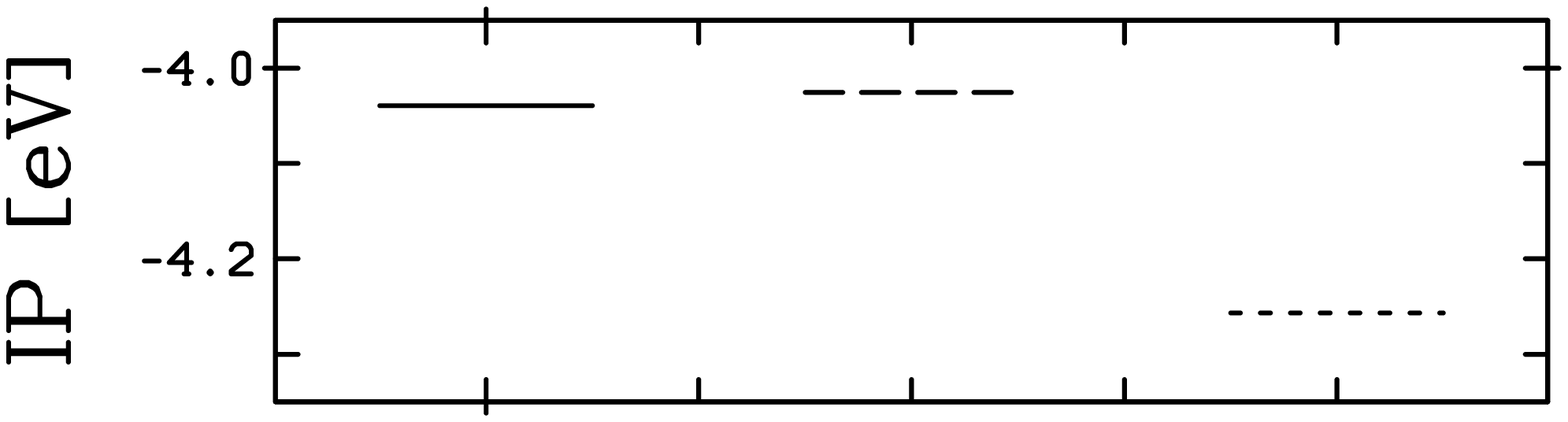,width=8.2cm}} 
\put(0,0){\epsfig{figure=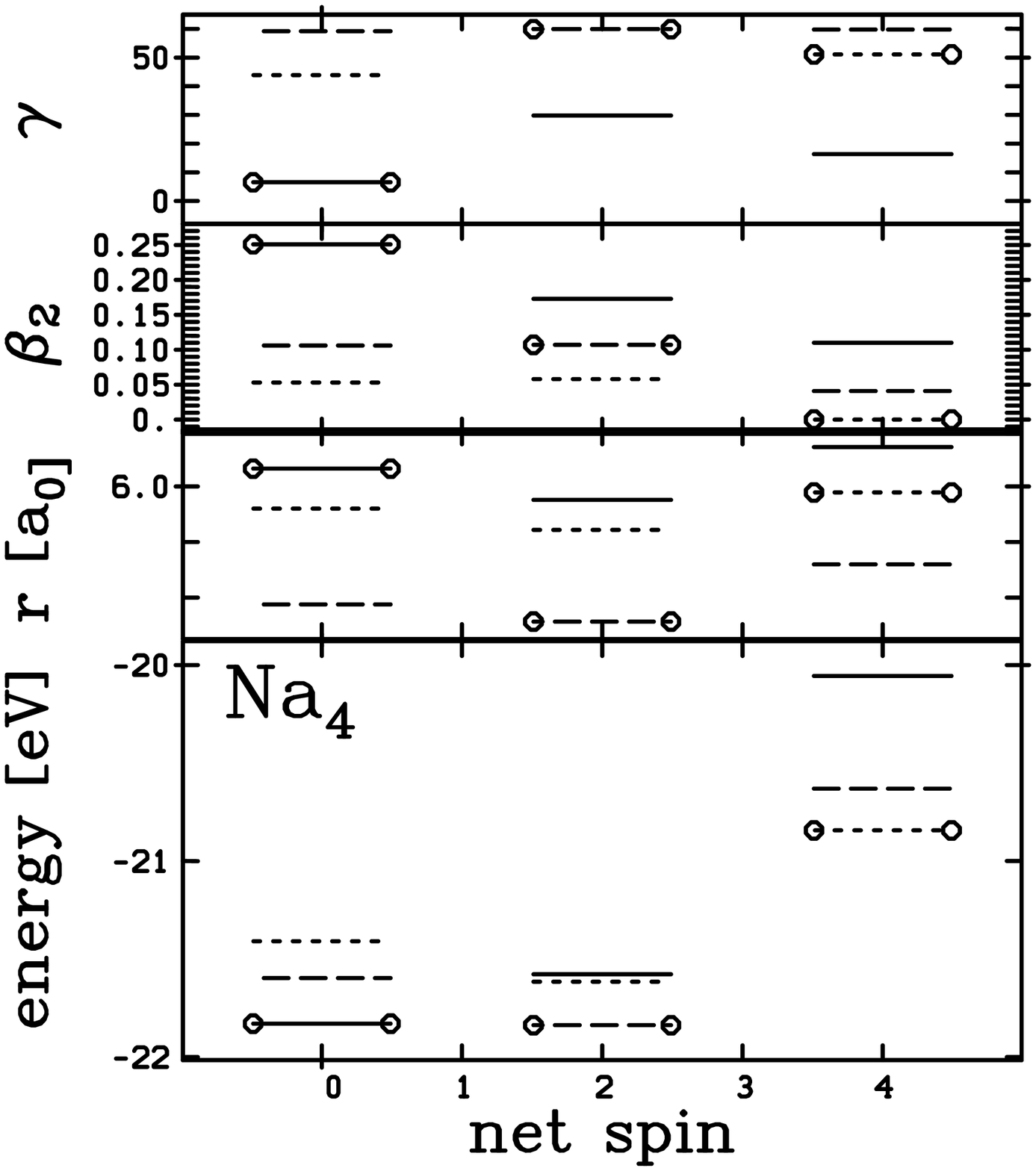,width=8.2cm}} 
\end{picture} 
\end{center} 
\caption{\label{fig:na4_detail}  
As figure \ref{fig:na3_detail} for the cluster Na$_4$. 
The assignment of line types to ionic 
configuration are: full line $\leftrightarrow$  optimized for spin=0,  
dashed $\leftrightarrow$ optimized for spin=2,  
dotted $\leftrightarrow$ optimized for spin=4. 
} 
\end{figure}

\begin{figure} 
\begin{center} 
\begin{picture}(164,290)(0,0) 
\put(27,228){\epsfig{figure=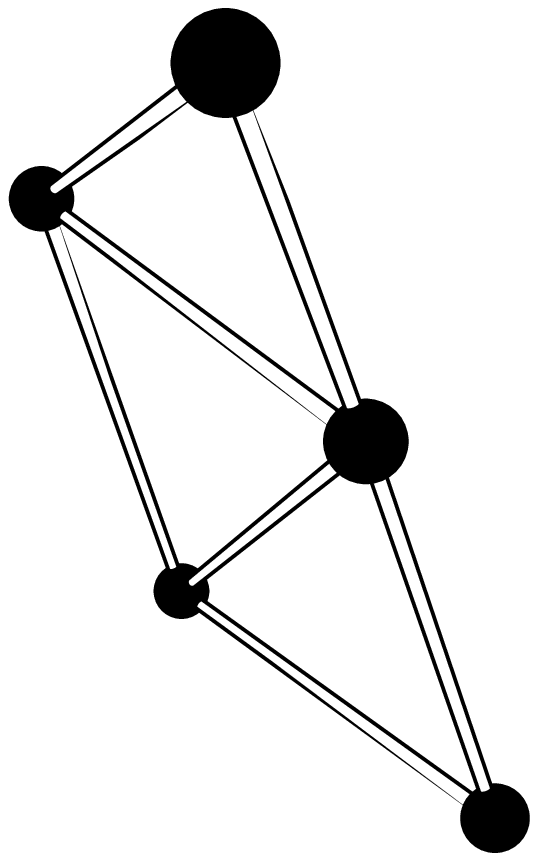,width=45pt}} 
\put(78,228){\epsfig{figure=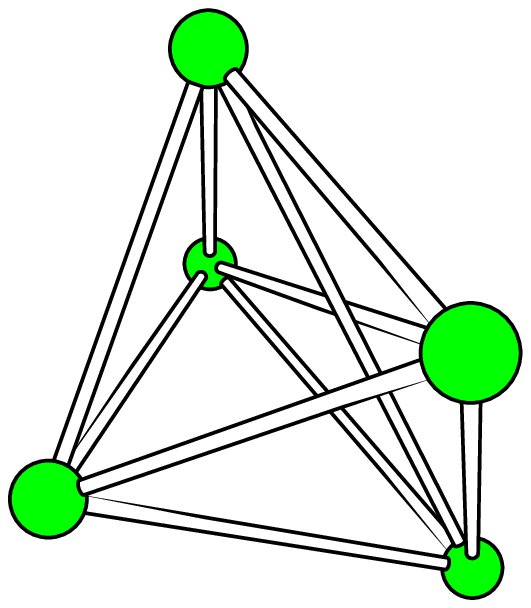,width=43pt}} 
\put(118,228){\epsfig{figure=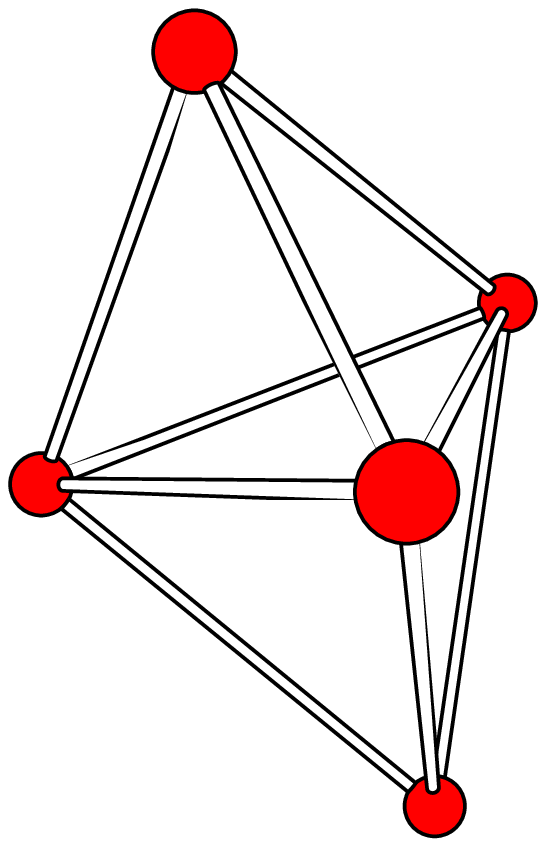,width=42pt}} 
\put(0,182){\epsfig{figure=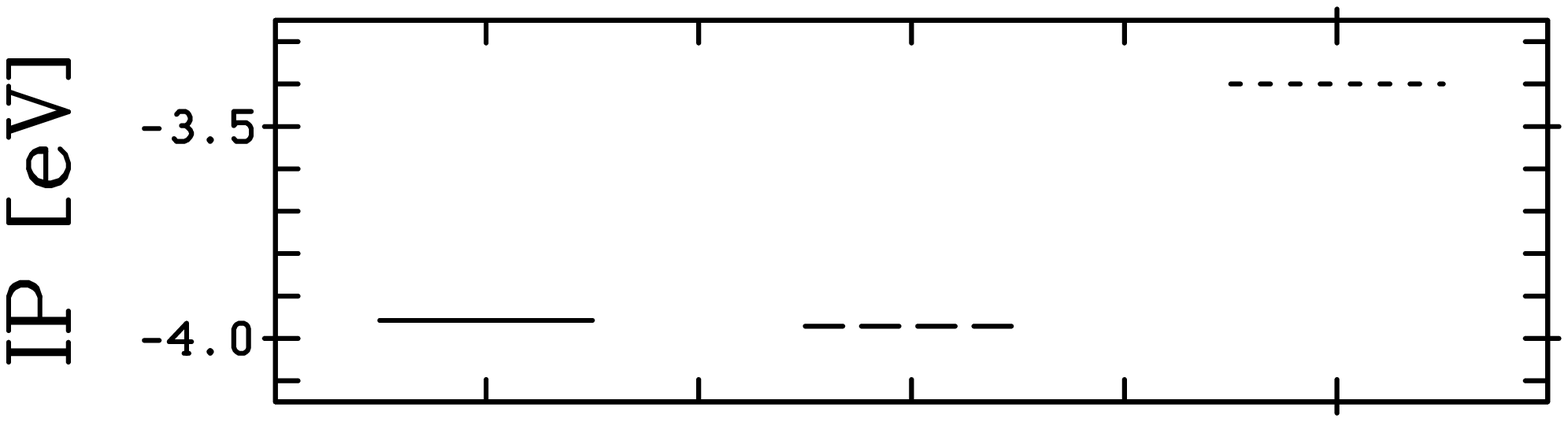,width=8.2cm}} 
\put(0,0){\epsfig{figure=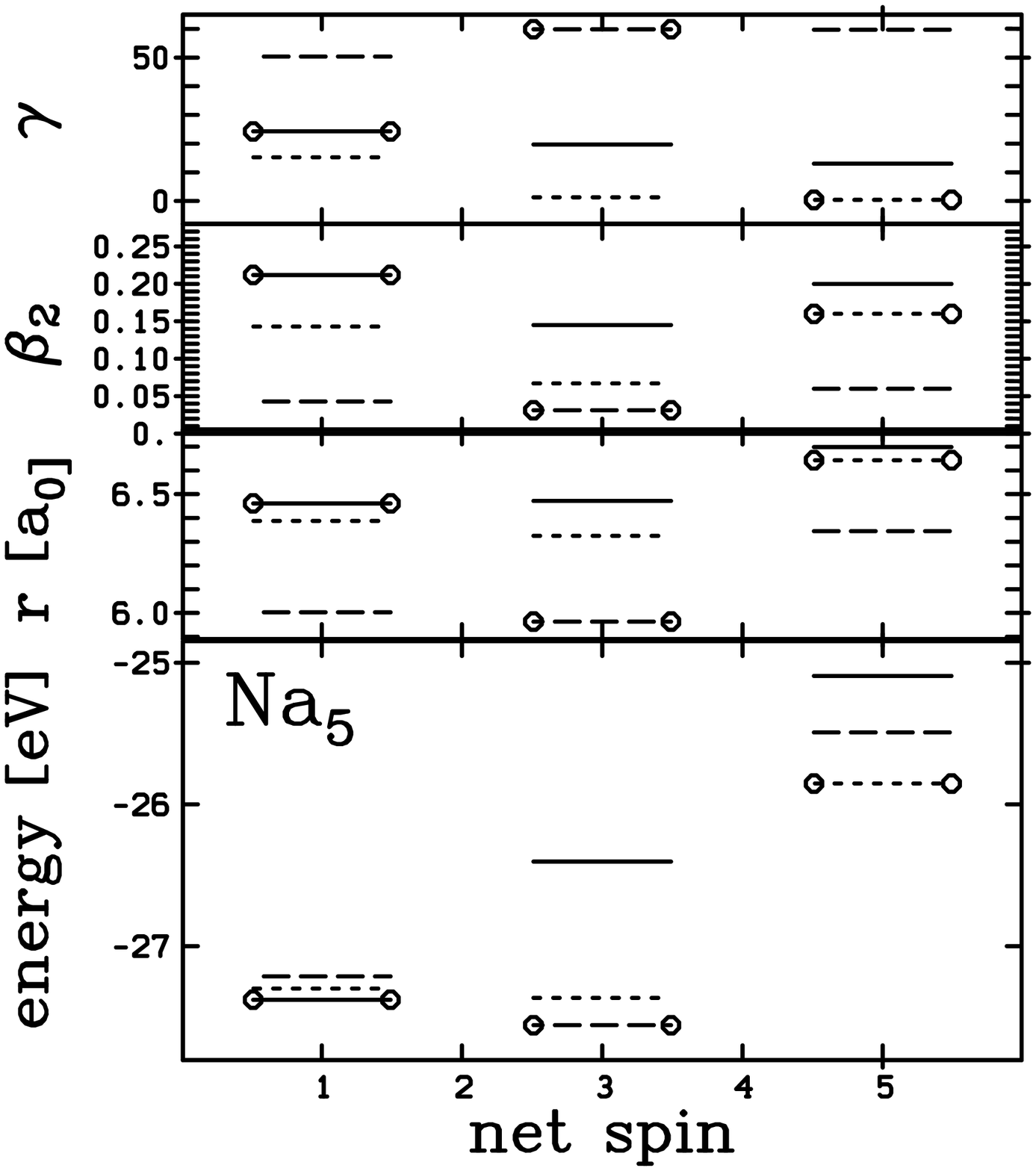,width=8.2cm}} 
\end{picture} 
\end{center} 
\caption{\label{fig:na5_detail}  
As figure \ref{fig:na3_detail} for the cluster Na$_5$. 
The assignment of line types to ionic 
configurations are: full line $\leftrightarrow$  optimized for spin=1,  
dashed $\leftrightarrow$ optimized for spin=3,  
dotted $\leftrightarrow$ optimized for spin=5. 
} 
\end{figure}

\begin{figure} 
\begin{center} 
\begin{picture}(164,290)(0,00) 
\put(24,228){\epsfig{figure=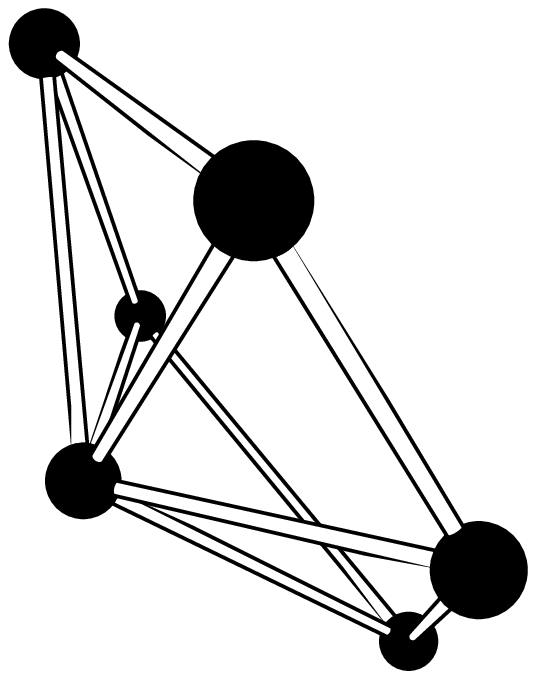,width=40pt}} 
\put(60,233){\epsfig{figure=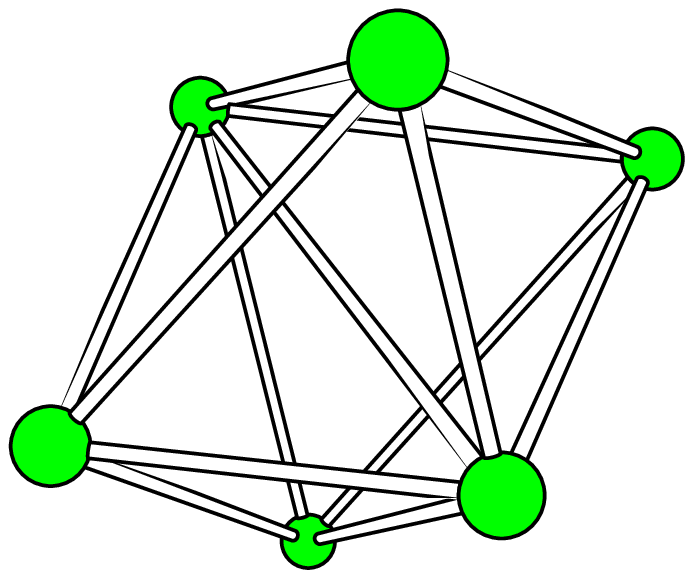,width=42pt}} 
\put(93,236){\epsfig{figure=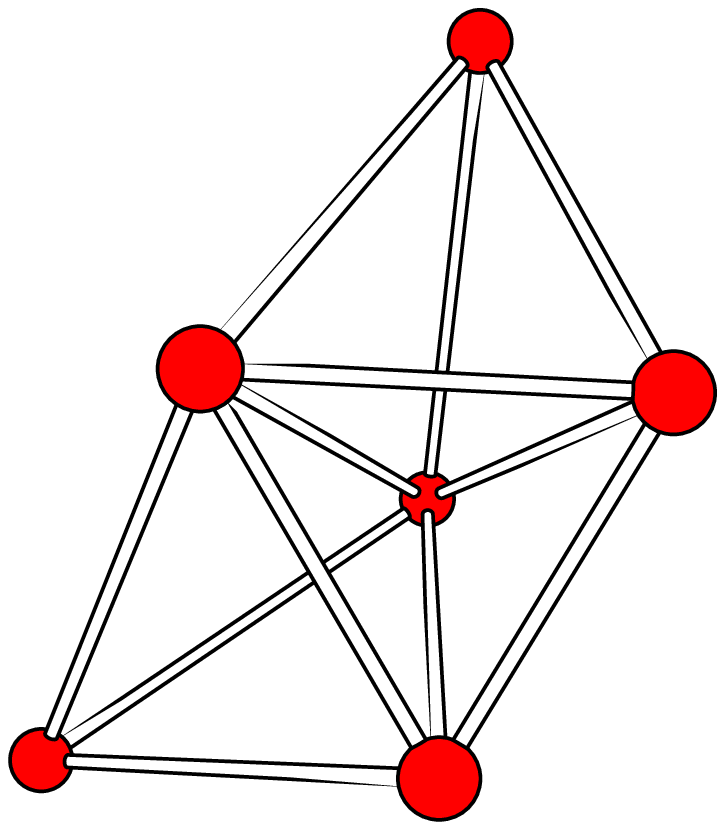,width=42pt}} 
\put(125,238){\epsfig{figure=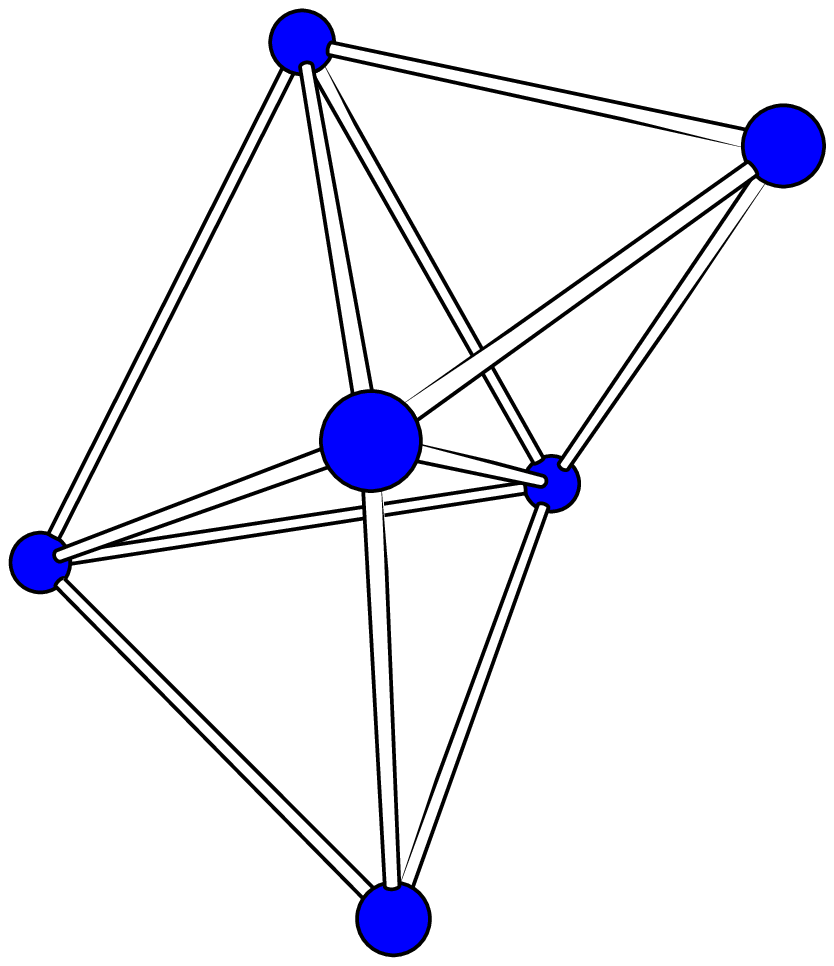,width=52pt}} 
\put(0,182){\epsfig{figure=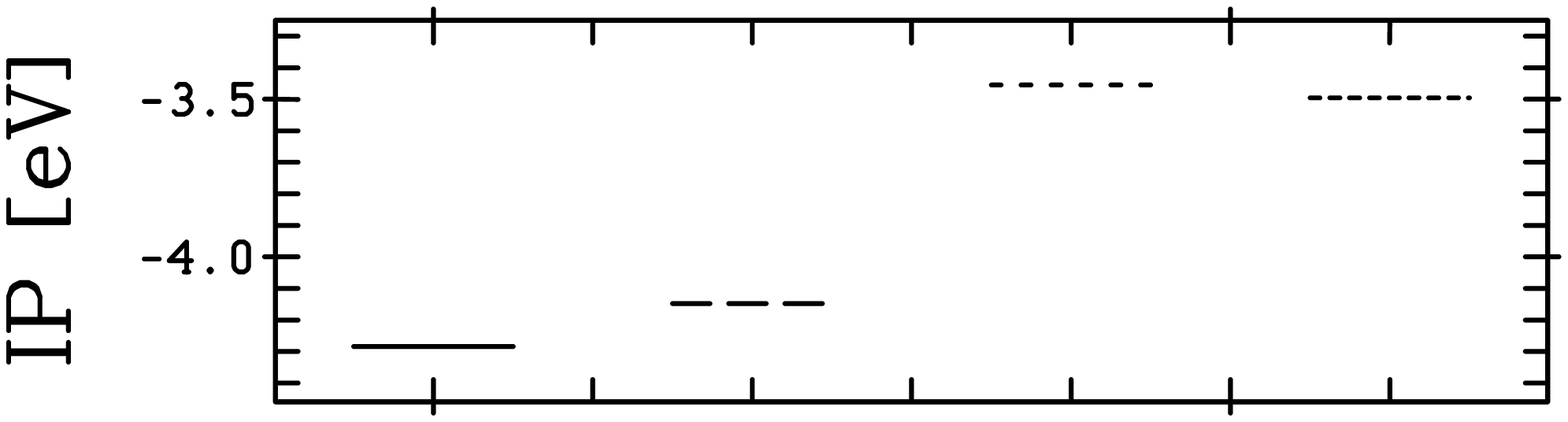,width=8.2cm}} 
\put(0,0){\epsfig{figure=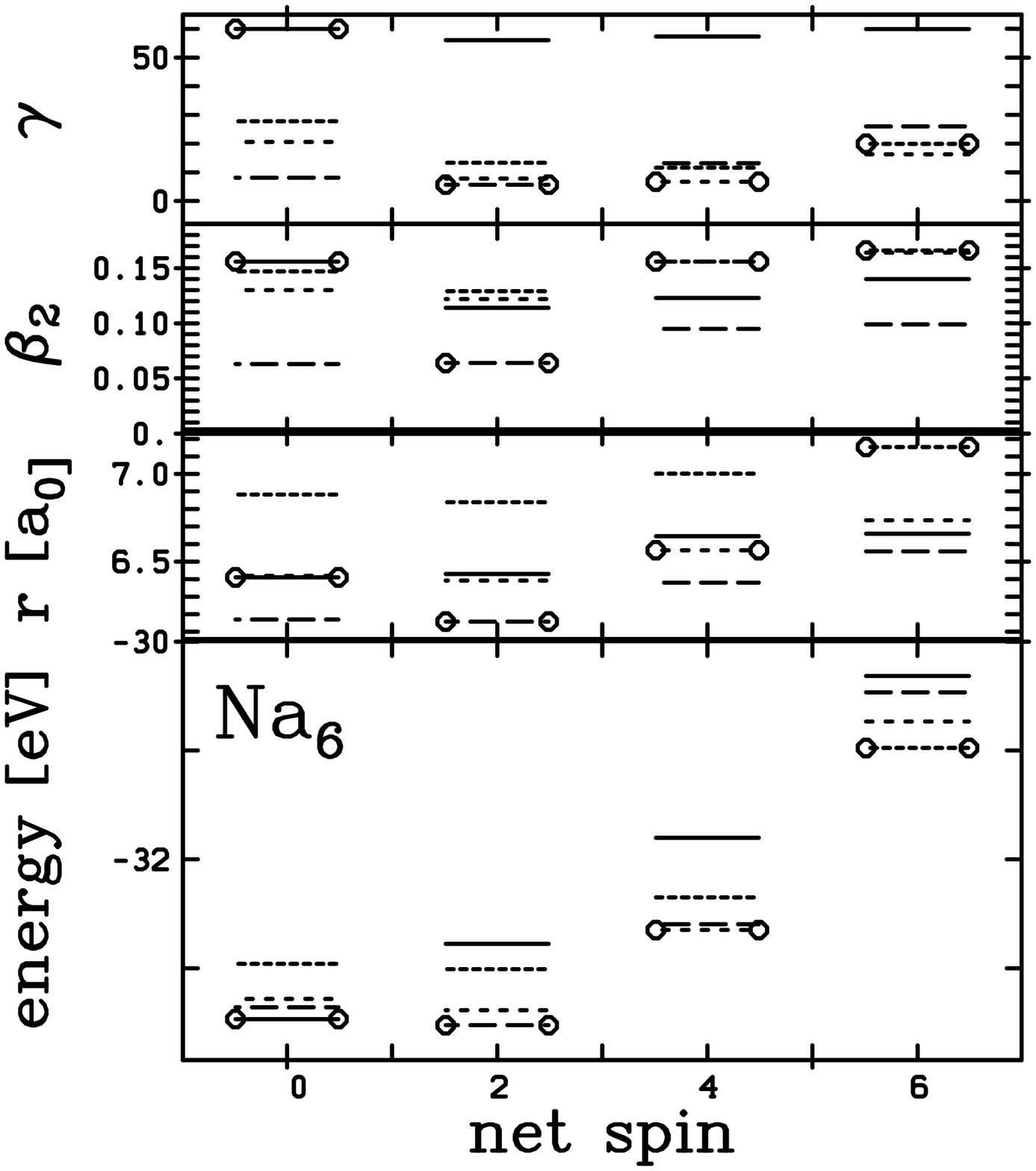,width=8.2cm}} 
\end{picture} 
\end{center} 
\caption{\label{fig:na6_detail}  
As figure \ref{fig:na3_detail} for the cluster Na$_6$. 
The assignment of line types to ionic 
configurations are: full line $\leftrightarrow$  optimized for spin=0,  
dashed $\leftrightarrow$ optimized for spin=2,  
dotted $\leftrightarrow$ optimized for spin=4, 
fine-dotted $\leftrightarrow$ optimized for spin=6. 
} 
\end{figure}

\begin{figure} 
\begin{center} 
\begin{picture}(164,290)(0,0) 
\put(28,228){\epsfig{figure=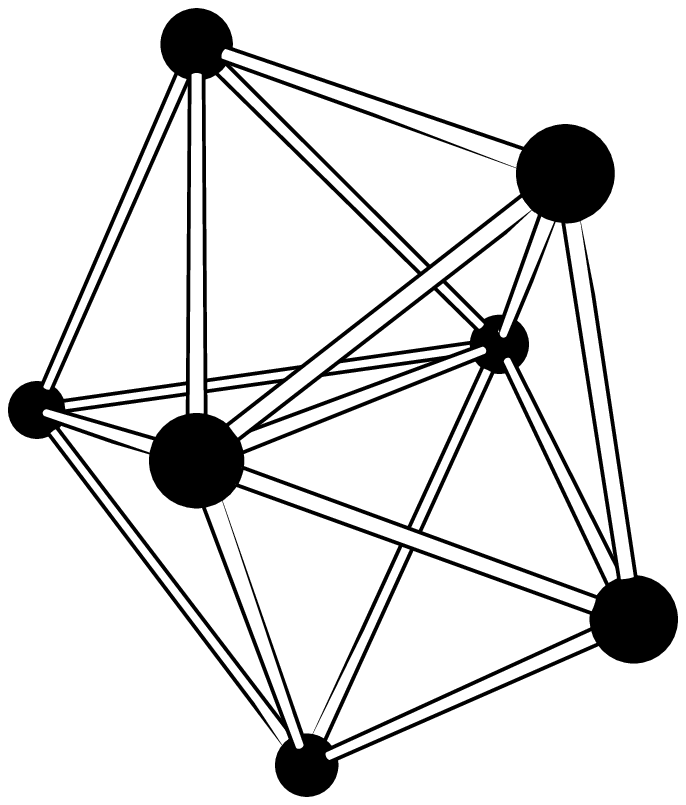,width=42pt}} 
\put(68,226){\epsfig{figure=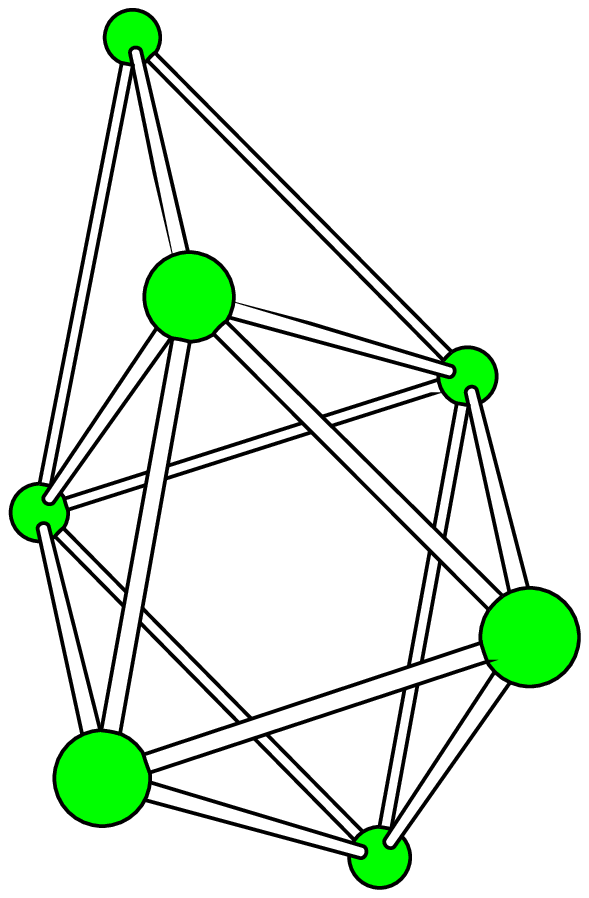,width=36pt}} 
\put(96,237){\epsfig{figure=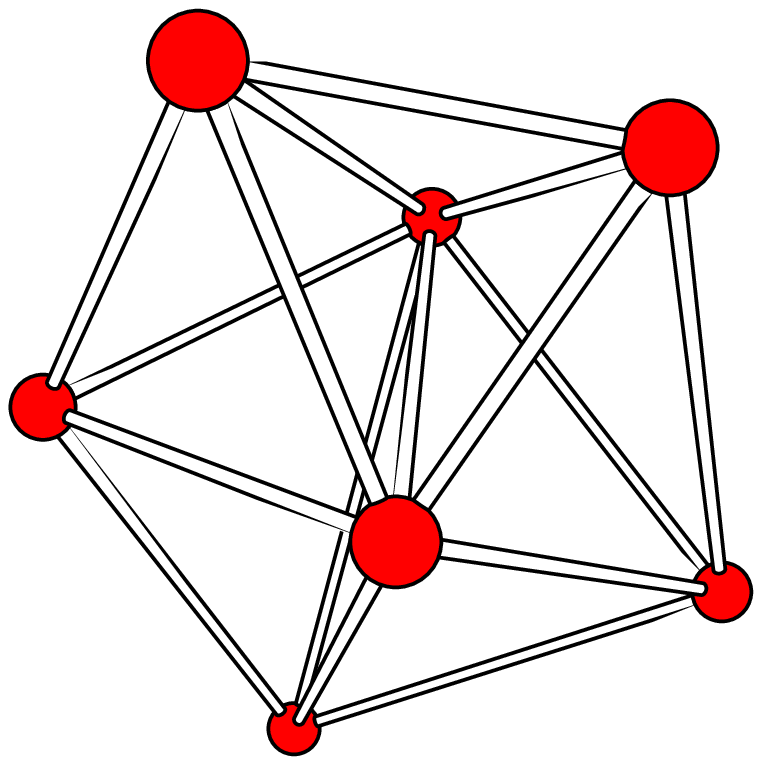,width=40pt,angle=90}} 
\put(129,227){\epsfig{figure=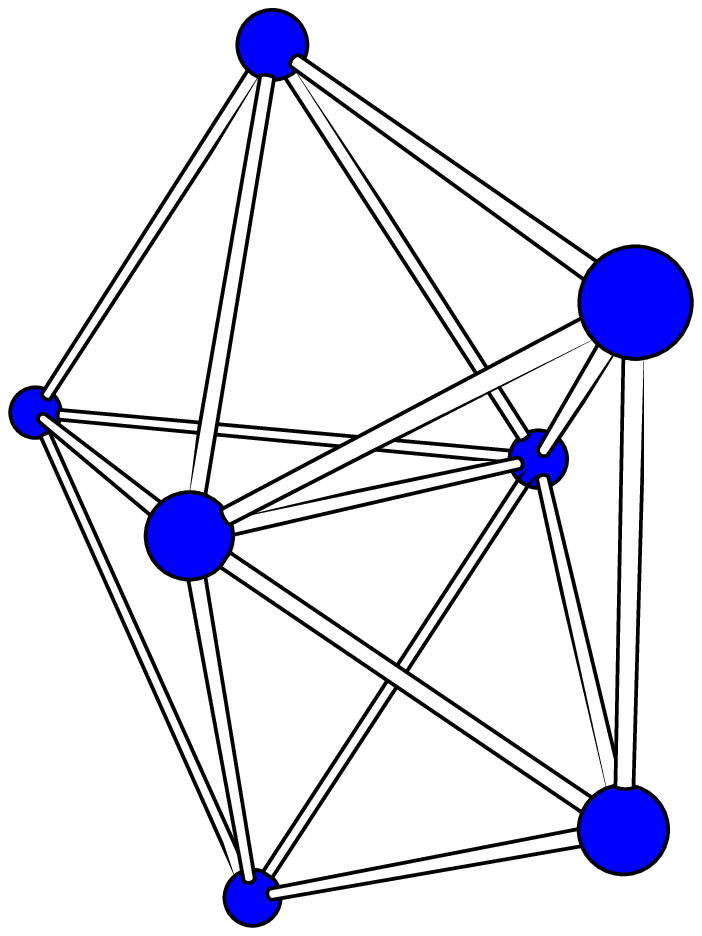,width=39pt}} 
\put(0,182){\epsfig{figure=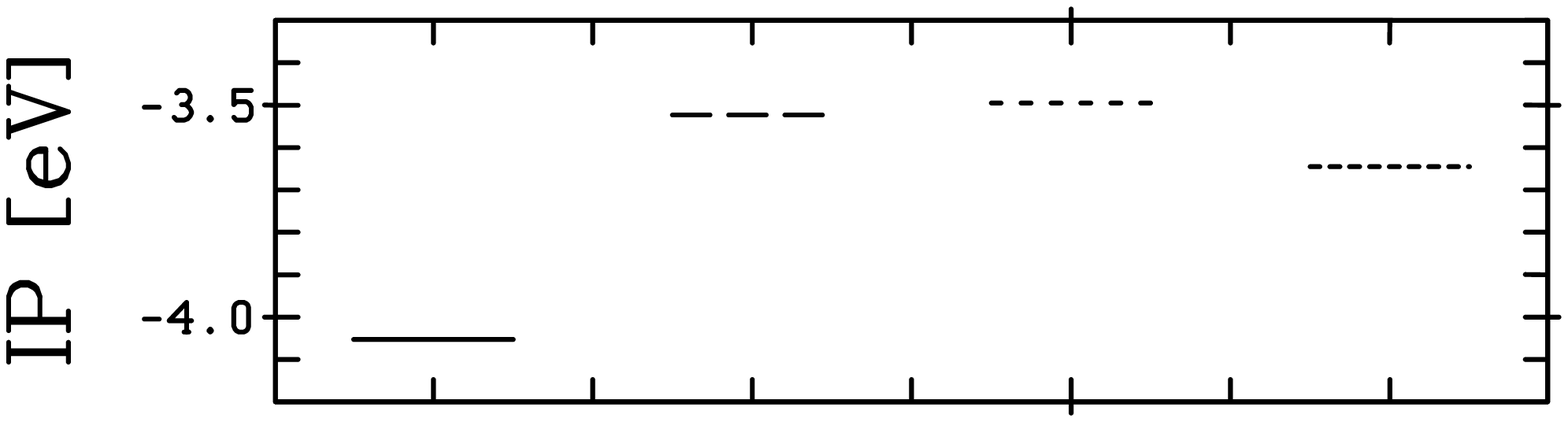,width=8.2cm}} 
\put(0,0){\epsfig{figure=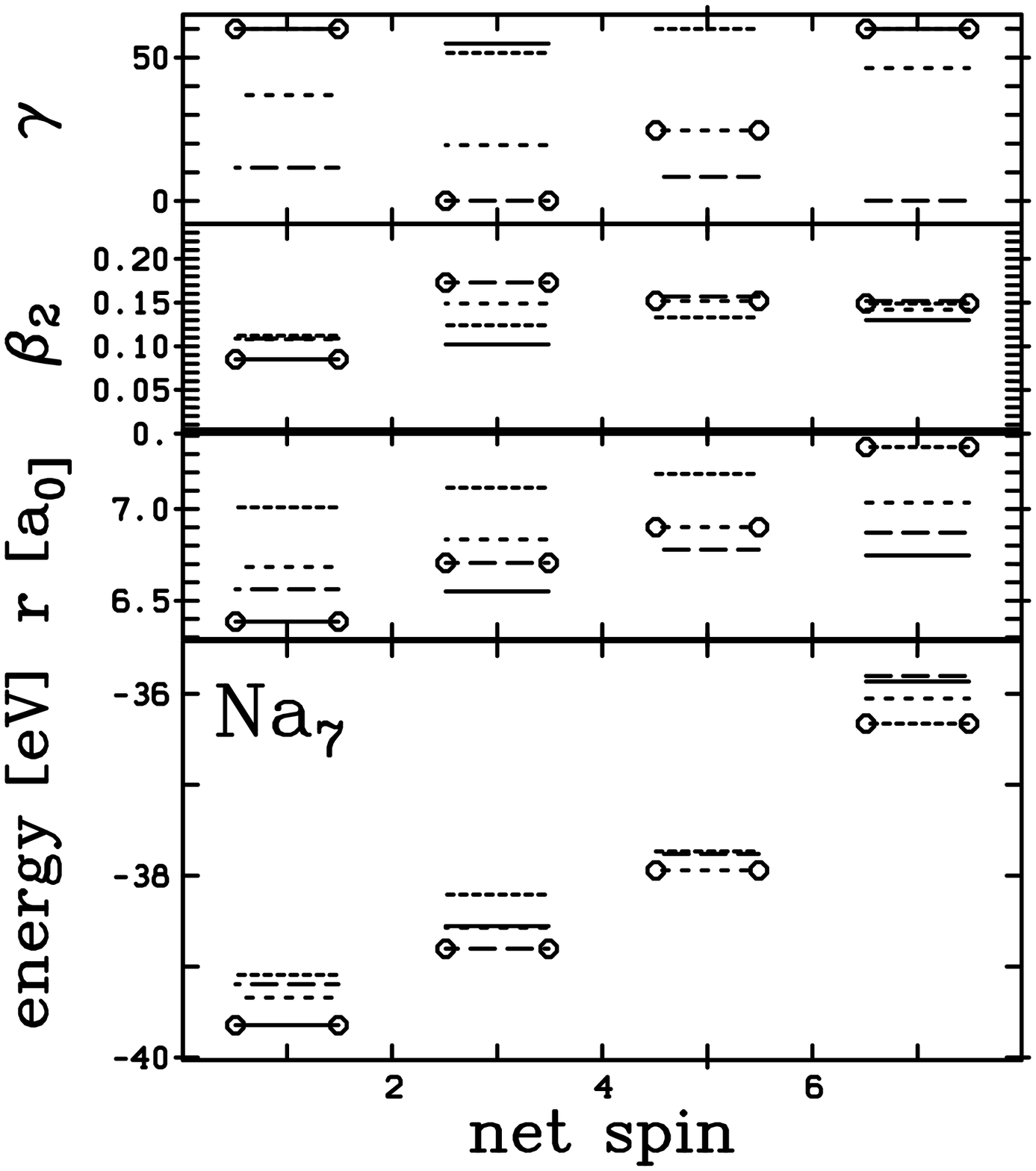,width=8.2cm}} 
\end{picture} 
\end{center} 
\caption{\label{fig:na7_detail}  
As figure \ref{fig:na3_detail} for the cluster Na$_7$. 
The assignment of line types to ionic 
configurations are: full line $\leftrightarrow$  optimized for spin=1,  
dashed $\leftrightarrow$ optimized for spin=3,  
dotted $\leftrightarrow$ optimized for spin=5, 
fine-dotted $\leftrightarrow$ optimized for spin=7. 
} 
\end{figure}

\begin{figure} 
\begin{center} 
\begin{picture}(164,290)(0,0) 
\put(28,228){\epsfig{figure=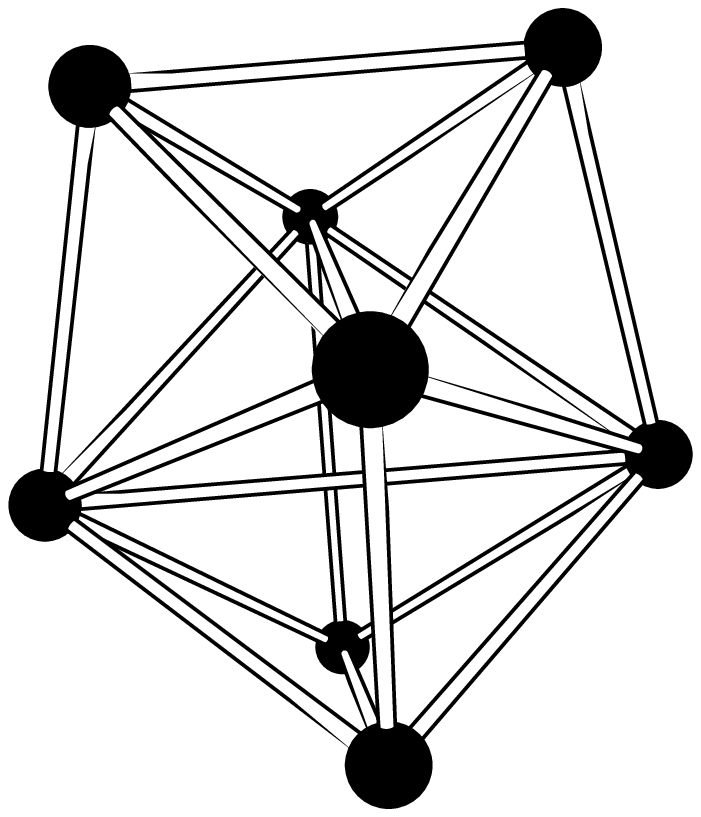,width=33pt}} 
\put(56,228){\epsfig{figure=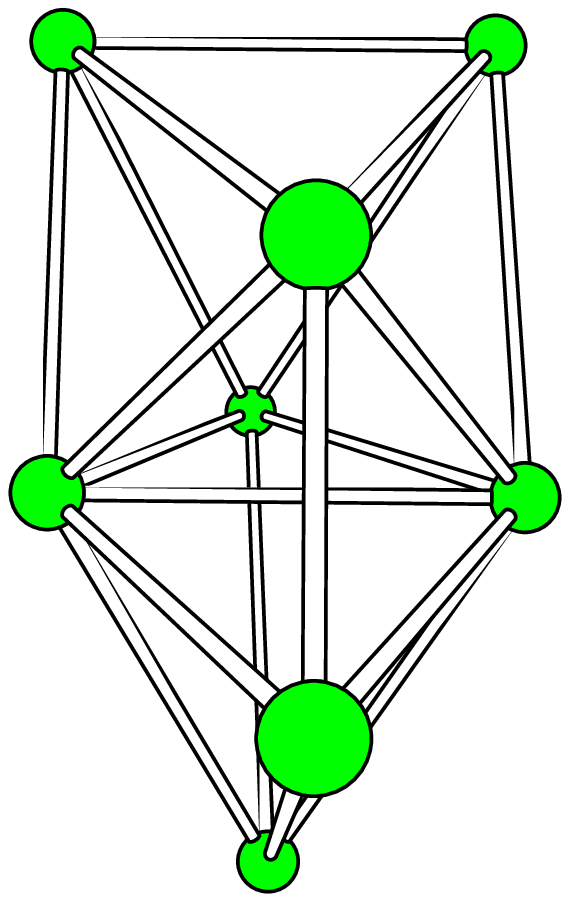,width=33pt}} 
\put(83,234){\epsfig{figure=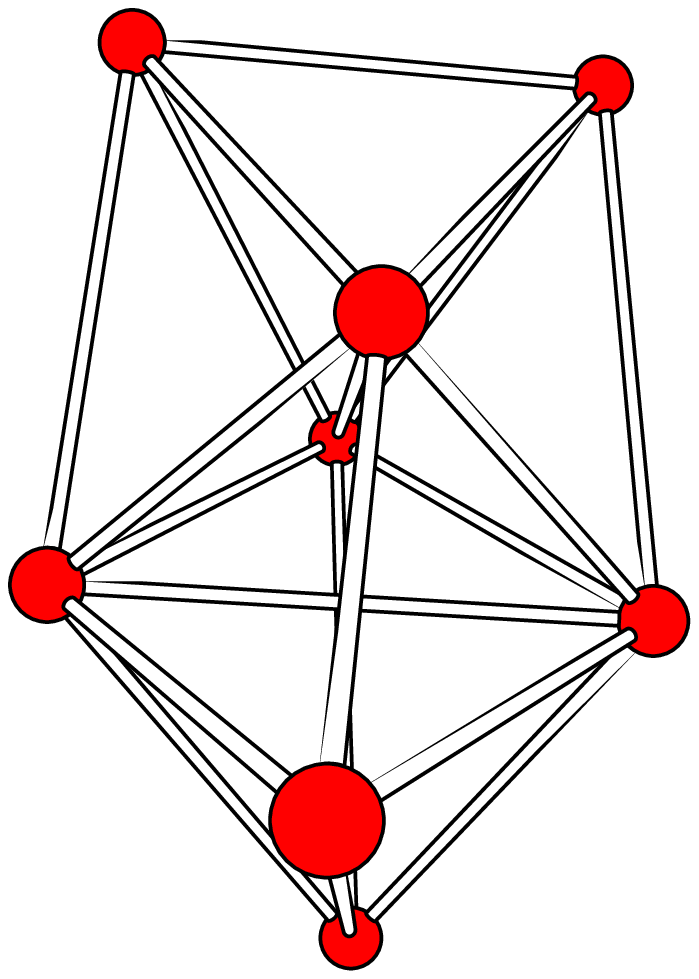,width=33pt}} 
\put(107,228){\epsfig{figure=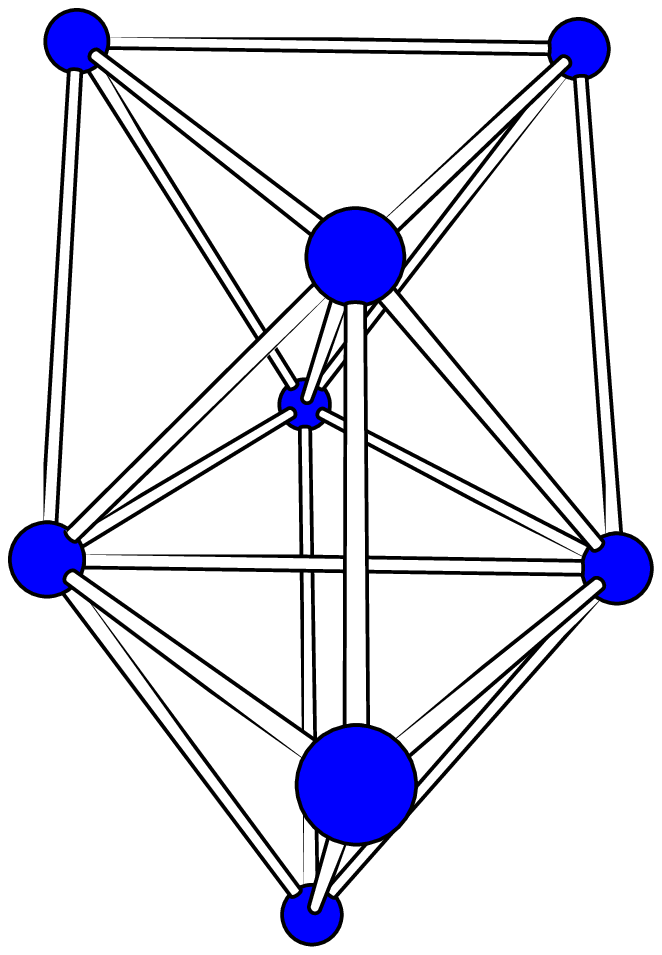,width=33pt}} 
\put(133,231){\epsfig{figure=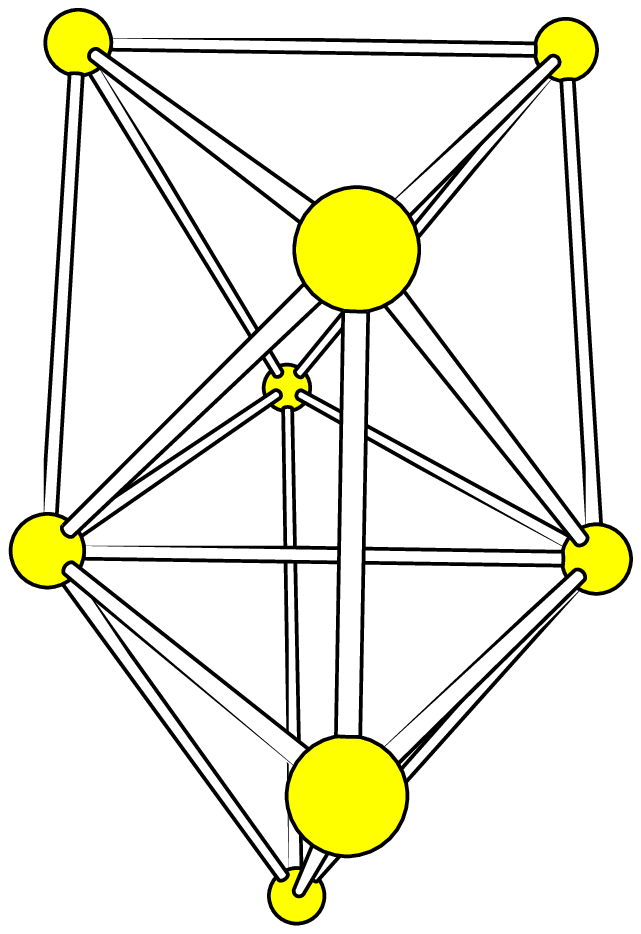,width=33pt}} 
\put(0,182){\epsfig{figure=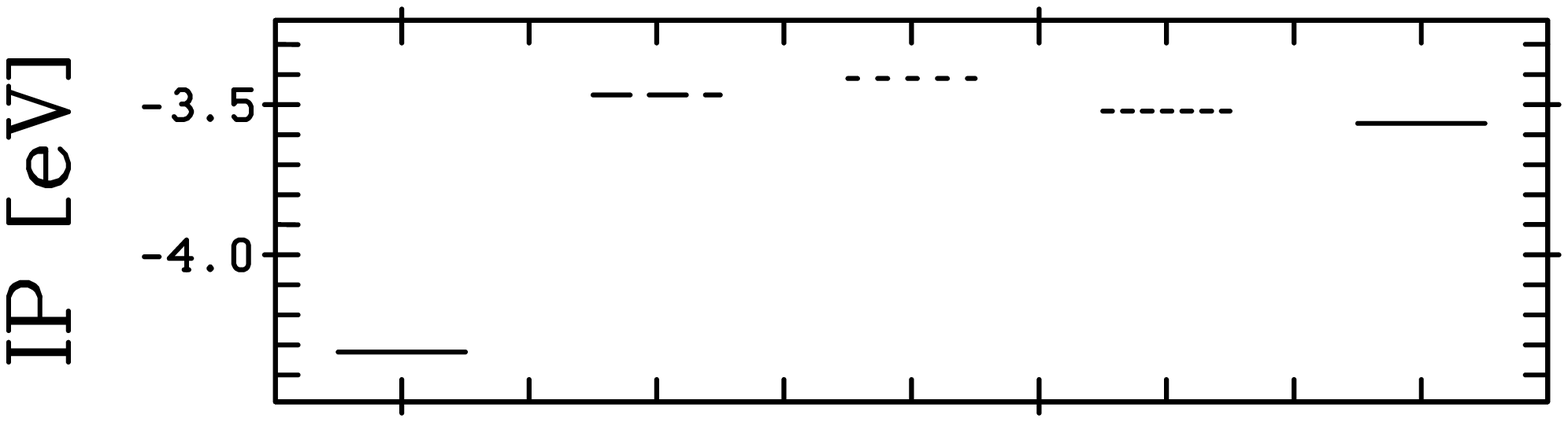,width=8.2cm}} 
\put(0,0){\epsfig{figure=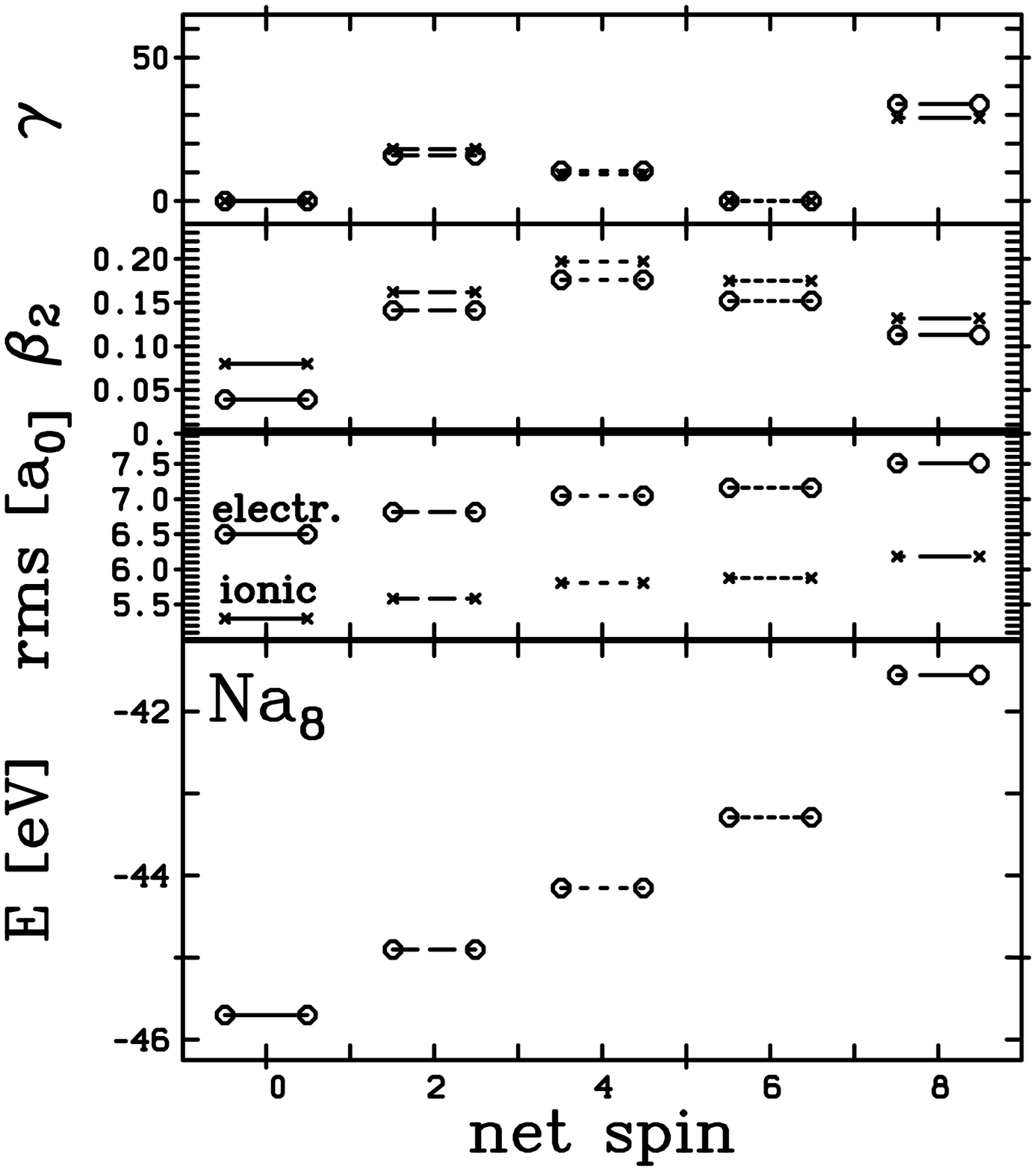,width=8.2cm}} 
\end{picture} 
\end{center} 
\caption{\label{fig:na8_detail}  
Global properties of Na$_8$ in various 
spin states: Lower panel = binding energy, second lower panel = 
r.m.s. radius, second upper panel = total quadrupole deformation, 
upper panel = triaxiality. The shape parameters are shown for the 
electronic distribution (lines embraced by circles) 
as well as for the ionic configuration (lines embraced by stars). 
They are defined in eq. (\ref{eq:shape}).   
} 
\end{figure}

\begin{figure} 
\begin{center} 
\begin{picture}(164,290)(0,0) 
\put(20,243){\epsfig{figure=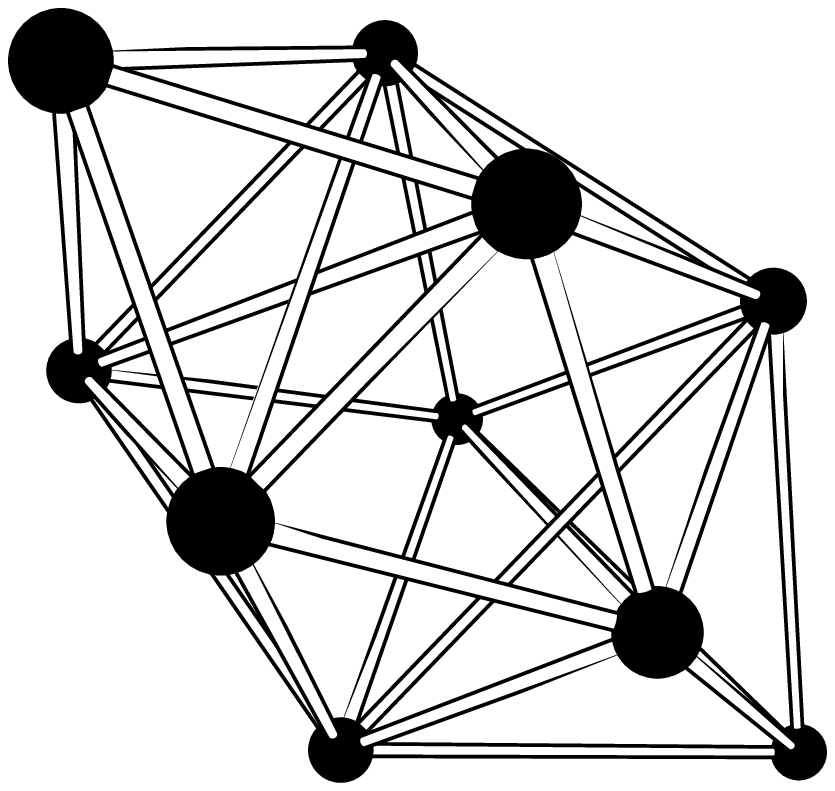,width=31pt,angle=-40}} 
\put(42,241){\epsfig{figure=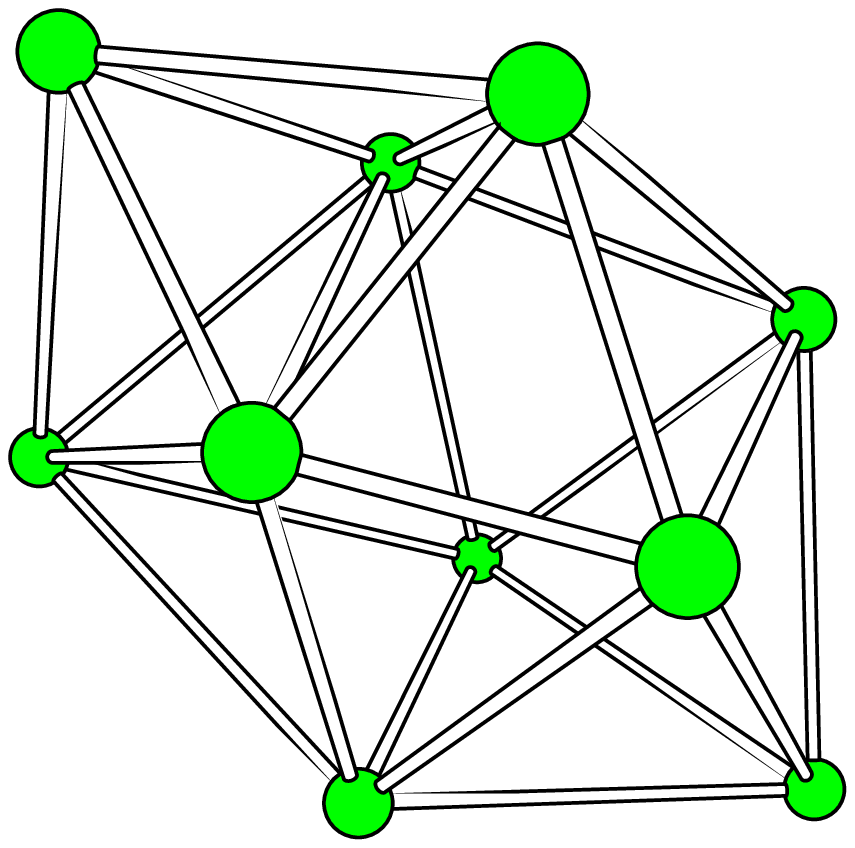,width=31pt,angle=-40}} 
\put(66,243){\epsfig{figure=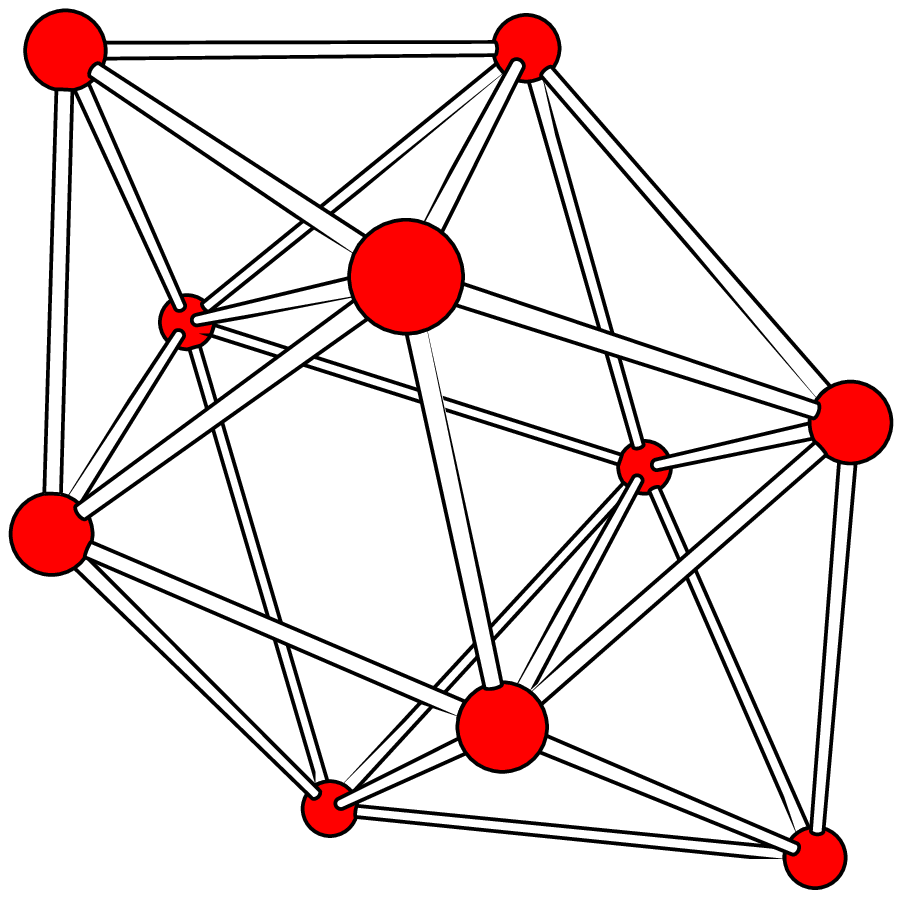,width=31pt,angle=-40}} 
\put(89,241){\epsfig{figure=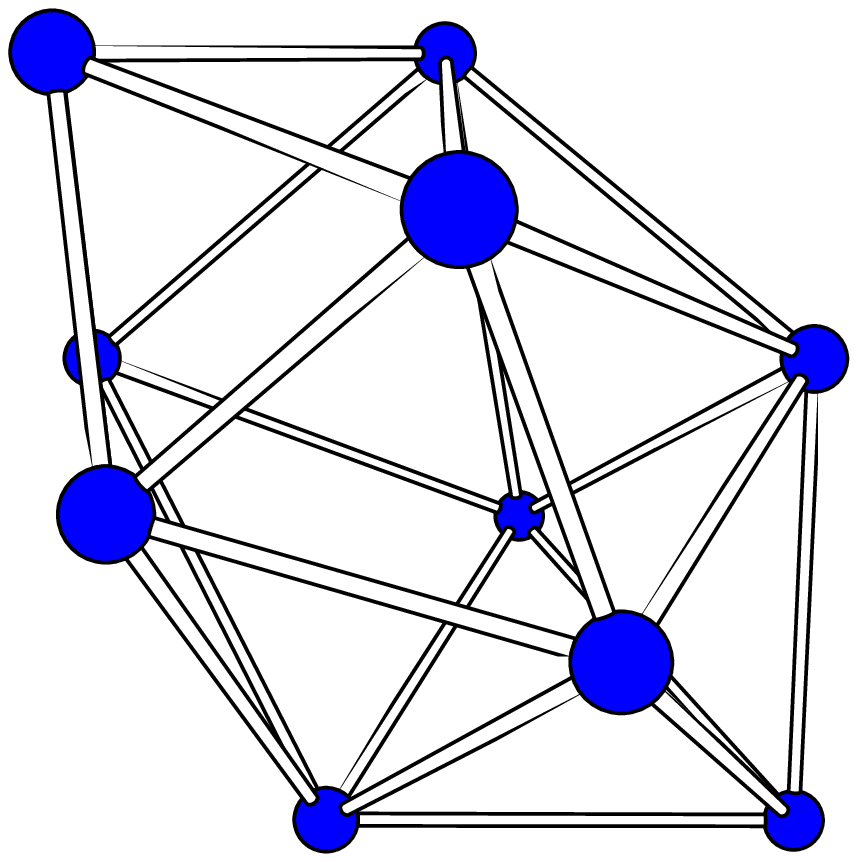,width=31pt,angle=-40}} 
\put(110,243){\epsfig{figure=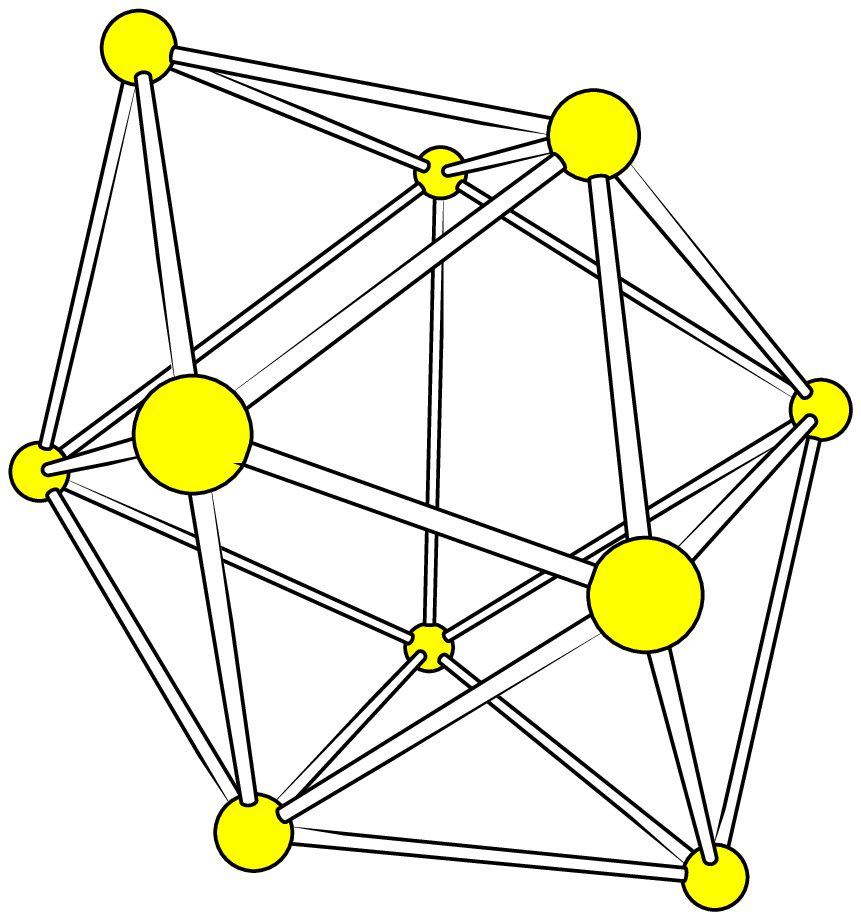,width=31pt,angle=-40}} 
\put(132,244){\epsfig{figure=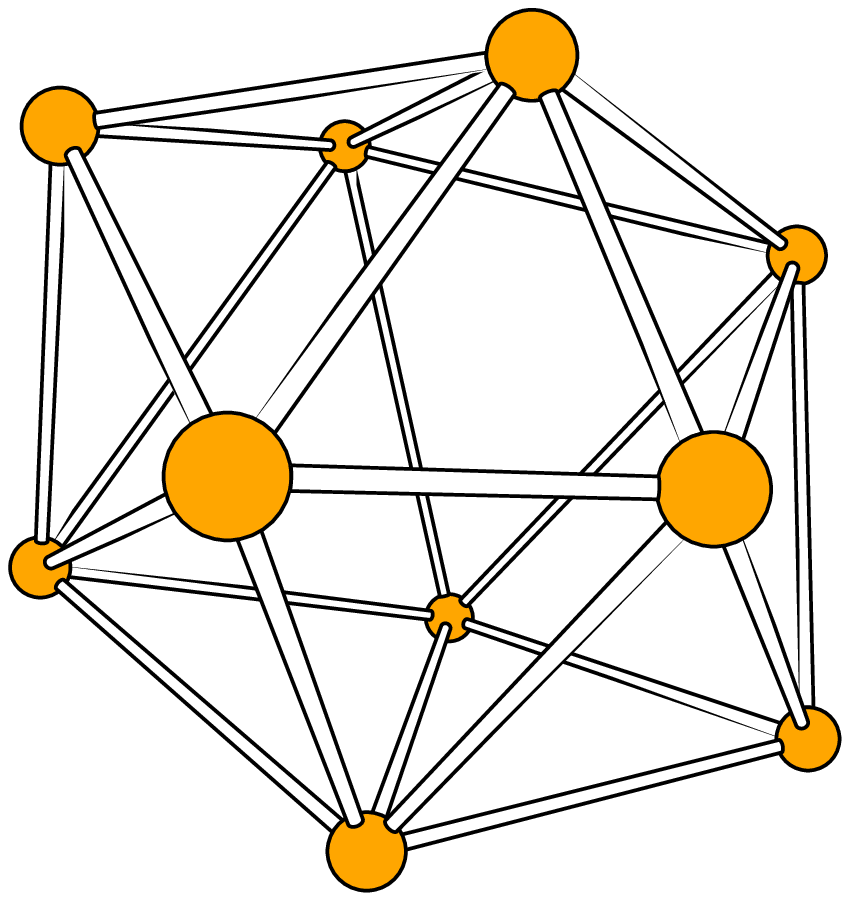,width=31pt,angle=-45}} 
\put(0,182){\epsfig{figure=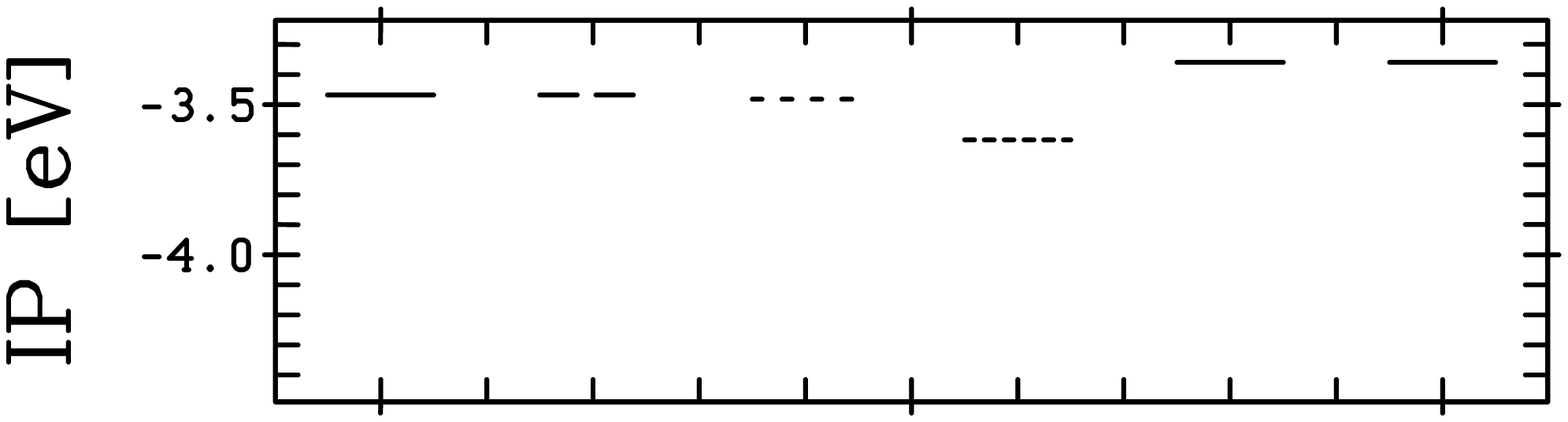,width=8.2cm}} 
\put(0,0){\epsfig{figure=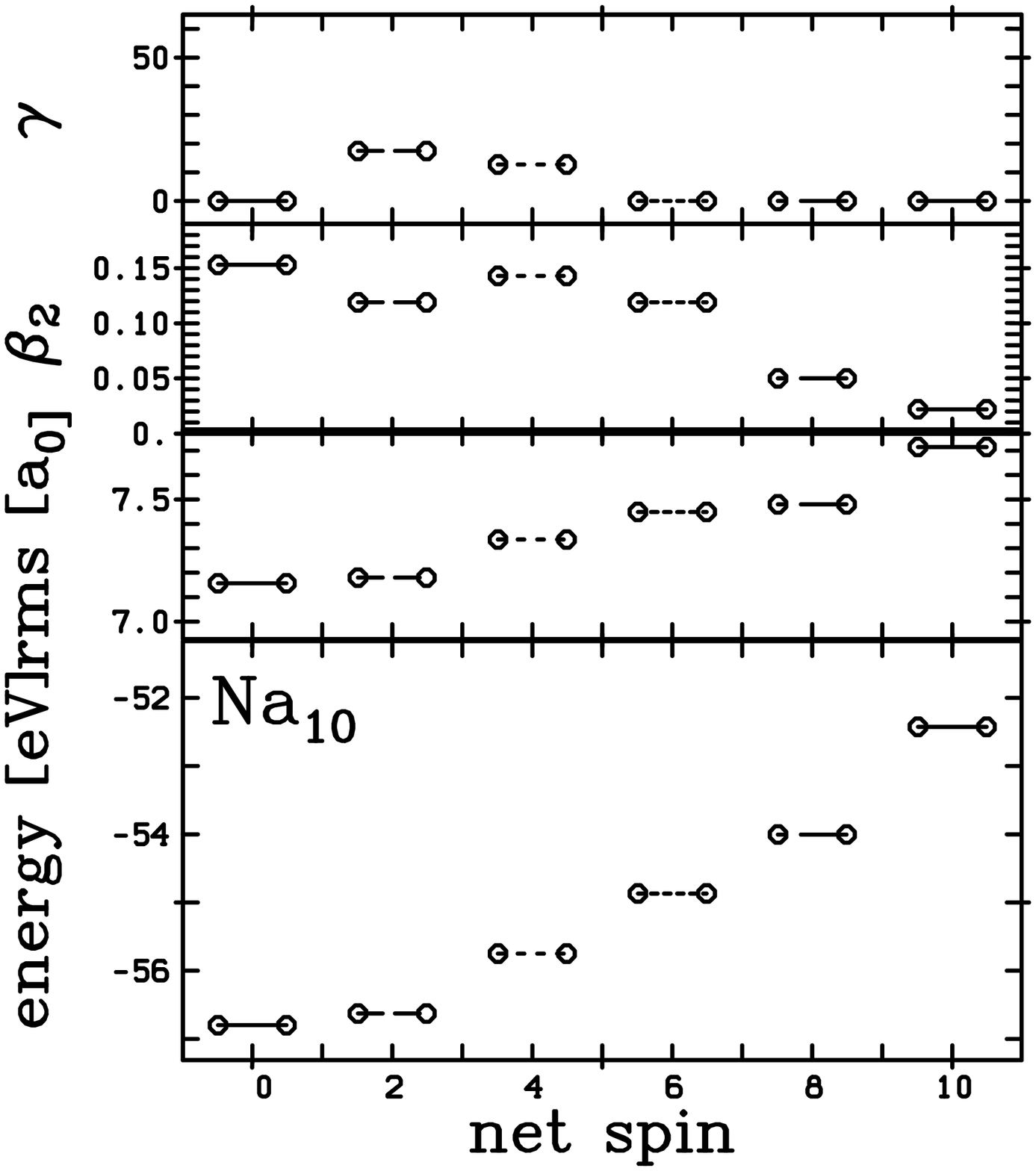,width=8.2cm}} 
\end{picture} 
\end{center} 
\caption{\label{fig:na10_detail}  
As figure \ref{fig:na8_detail} for the cluster Na$_{10}$.} 
\end{figure}

Figures \ref{fig:na3_detail}--\ref{fig:na10_detail} show the global 
properties (cohesive energy, radius, deformation, IP) of all 
considered clusters together with a graphical illustration for the 
ionic structures. These figures are all built according to the same 
scheme and we discuss them in one stroke. Looking at the binding 
energies, we see (with few exceptions) that the energies for different 
ionic configurations but same net spin gather in densely packed small 
blocks while large energy shifts emerge when changing electronic net 
spin.  The binding energies thus depend predominantly on the 
electronic net spin while details of the ionic configuration make less 
effect, although there can be substantial ionic rearranegements, 
particularly for the smaller clusters. Questions of spin stability may 
thus be discussed in a first step on grounds of electronic dynamics 
(as done to some extend in section \ref{sec:spinstab}). The ionic 
rearrangement follows in a second step. This view is not only 
suggested by energetic considerations but also for reasons of time 
scale. Electronic transitions run at the order of a few fs while ionic 
motion takes hundreths of fs. 
 
The radii basically follow the trends of the binding energy, deeper 
binding relates to smaller radii. This holds strictly for the cases 
where electrons and ionic configuration are simultaneously optimized 
(bars distinguished by circles in the figures). The other combinations 
of electron spin and ionic configurations show variations which are 
larger than the respective variations in energy and which can, in 
contrast to the energy, vary in both directions around the optimized 
result. The deformation parameters $\beta$ and $\gamma$ also show 
large variations with ionic configuration at given net spin. These 
variations are as large, often even larger, as the changes with net 
spin. This happens because the electrons like to follow the given 
ionic shape in order to minimize the Coulomb energy \cite{CAPS}.   
 
The relation between electronic and ionic shape is visible in figure 
\ref{fig:na8_detail}. It shows the shape parameters for both species 
(ionic with stars, electronic with circles). Ionic and electronic 
radii follow precisely the same trend. But the electronic radius is 
systematically larger than the ionic radius. This is due to the much 
smoother electronic surface distribution. The deformation parameters 
$\beta$ and $\gamma$ coincide for electrons and ions. This 
demonstrates that the electrons follow the ionic distribution or vice 
versa.  It becomes fully obvious when looking at the $\beta$ and 
$\gamma$ across the various spins and comparing results only for the 
same line type which means the same ionic configuration. In this way, 
the results show much less variation. 
 
Let us now concentrate on the simultaneously optimized configurations 
(indicated by circles in the figures), i.e. the electronic plus ionic 
ground states at given spin, and let us go through the examples top 
down, i.e. from the largest sample to the smallest. Figure 
\ref{fig:na10_detail} for Na$_{10}$ shows the expected trends.  The 
unpolarized configuration is the preferred one and there is a steadily 
increasing energy towards the fully polarized case.  Two details go a 
bit against these general trends. First, the case with net spin 2 has 
very small energy difference to the spin 0 ground state. This 
configuration consists out of 6 spin-up with 4 spin-down electrons and 
the 4 electrons of one spin constitute a magic electron shell. The 
slight ``magicity'' can also be read off from the dip of the 
deformation $\beta$ at spin 2. The second detail concerns the state 
with spin 10. The very small $\beta$ there indicates a nearly 
spherical shape. Again we meet a magic number where 10 electrons of 
one spin form a closed shell. In this case, the ``magicity'' is not 
strong enough to be honored by the binding energy. But it suffices to 
drive the spherical shape.

The results for Na$_8$ in figure \ref{fig:na8_detail} come even closer 
to what one would have naively expected. There is an almost 
equidistant rise in energy and radius with increasing spin.  But mind 
that we have here an enhanced preference of the unpolarized state 
by a fully developed magic electron number at spin 0, namely 4 
electrons spin-up and 4 spin-down. This is corroborated by the fact 
that this system again has low $\beta$ and is thus close to spherical 
shape.  The case of Na$_7$ in figure \ref{fig:na7_detail} is much 
similar to Na$_8$. No surprise, because the lowest spin 1 coincides 
again with the magic electron closure of the 4 spin-up electrons. 
Comparing the cases Na$_8$ and Na$_7$ with Na$_{10}$, we conclude that 
there is a general trend toward unpolarized systems, but that shell 
closures can change the picture in detail.  That means that small net 
spin may emerge as ground state configuration if one spin species has 
a magic electron number and the other species just suffices to 
compensate to net spin 2, or 3 respectively.  This effect will 
become more obvious for the smaller systems.

The example of Na$_6$ shown in figure \ref{fig:na6_detail} is the
first case where shell closure is compensating the trend to zero
polarization. The global ground state happens to be configuration with
spin 2, not surprisingly just the case covering a magic shell of 4
spin-up electrons. The radius follows the trend of the energy, and
last not least, the ``magicity'' is again indicated by a drop in
deformation $\beta$. The situation is similar again in the next lower
Na$_5$ cluster. The ground state is here with spin 3 because this
contains the magic shell of 4 spin-up electrons.  In both examples,
however, the notion of a ground state configuration has to be taken
with a grain of salt. The energy difference to the minimum spin state
is extremely small. A clear clut decision between ground state and
first isomer may be beyond the reliability of LSDA. In any case, the
qualitative result will persist, namely that the spontanously spin
polarized state is competitive with the minimum spin state. And this
has interesting consquences on the magnetic response as was discussed
in \cite{magnresp}.

The fully spin polarized Na$_4$ is again close to spherical in 
compliance with the shell closure at $N=4$ spin-up electrons. 
However, the energetically favored ground state configuration is 
${{\rm Na}_4}^{31}$ having net spin 2. The price for full spin 
polarization is here higher than the gain from shell closure. 
Finally, nothing peculiar can be seen for the smallest system 
Na$_3$. It is necessarily a flat object and thus has always a certain 
deformation. Note that triaxiality $\gamma$ is not so well defined 
and thus fluctuates for such a small system. 
 
The small clusters Na$_4$ and Na$_5$ show a strong dependence of the 
shape on spin. The unpolarized configurations are planar while the spin 
polarized states extend in three dimensions. This is due to the drive 
of the $N_{\rm spinup}=4$ shell to sphericity. The differences in 
shape become less dramatic with increasing $N$. But one still has for 
any $N$ the influence of magic $N_{\rm spinup}$ producing minima in 
$\beta$. The effect is obvious for the $N_{\rm spinup}=4$ shell.  But 
one sees it also on the $\beta$ from the softer $N_{\rm spinup}=10$ 
shell. There is a pronounced minimum for Na$_{10}$. For Na$_8$, the 
maximum $\beta$ is precisely in between 4 and 10 whereas the fully 
polarized Na$_8$ clearly shows the descend of $\beta$ toward the 
magic shell. 
 
Besides total energies, figures 
\ref{fig:na3_detail}--\ref{fig:na10_detail} provide also the 
ionization potentials (IP) of the various spin configurations.  The IP 
characterizes the stability of a system against removal of an electron. The 
magic shell $N_{\rm spinup}=4$ is clearly visible. The overall pattern show 
the typical stepping up above at $N_{\rm spinup}=4$ because the $1p$ state is 
fully occupied and the less well bound $1d$ state is going to become filled. 
The spin-saturated full-shell case $(N,N_{\rm spinup})=(8,4)$ is best 
bound. However, the differences to all others cases are very small.


A word is in place here about Hund's rules. These are formulated for 
atoms and they state that electrons in an open shell arrange 
themselves into a maximal spin-polarization to render the ground state 
non-degenerate. Clusters have an alternative, and more effective, way 
to arrange an unambiguous ground state, the Jahn-Teller effect, 
i.e. they drive into a deformation for which the electron occupation 
is unique. The two mechanisms compete and the Jahn-Teller deformation 
usually wins. Still, there are occasionally spin polarized isomers in 
clusters with truly triaxial shapes \cite{magnresp,spinjel}. These 
previous studies considered only small polarization up to net spin 
two. The present investigation goes up to any spin and reveals that 
shell closures for spin-polarized systems add an extra preference for 
electron number 4 (and 10) in one spin species. The extra binding 
leads to spin polarized ground states in very small Na clusters. It is 
so to say Hund's rules enhanced by shell effects.

\begin{figure} 
\centerline{\epsfig{figure=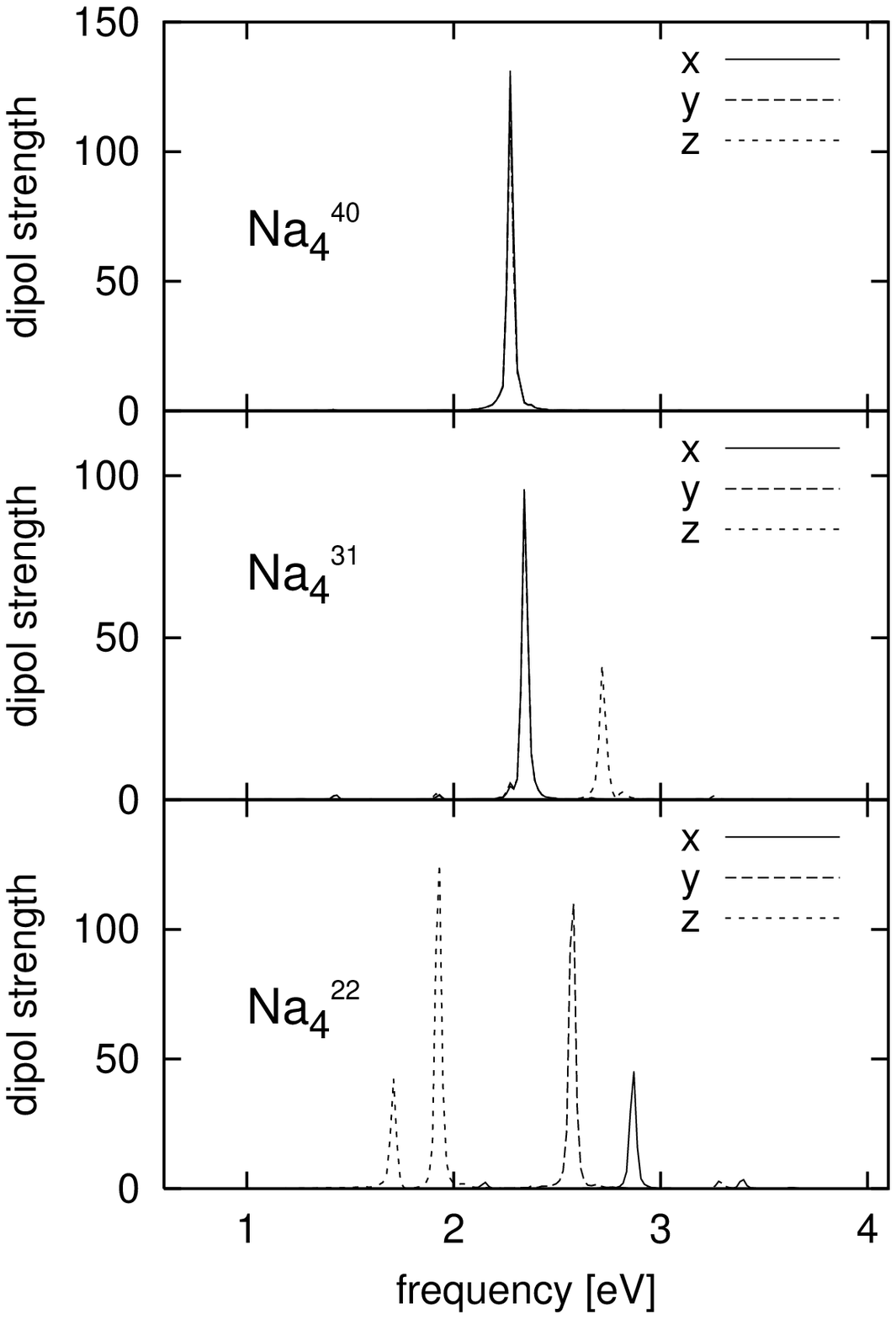,width=7.8cm}\;
\epsfig{figure=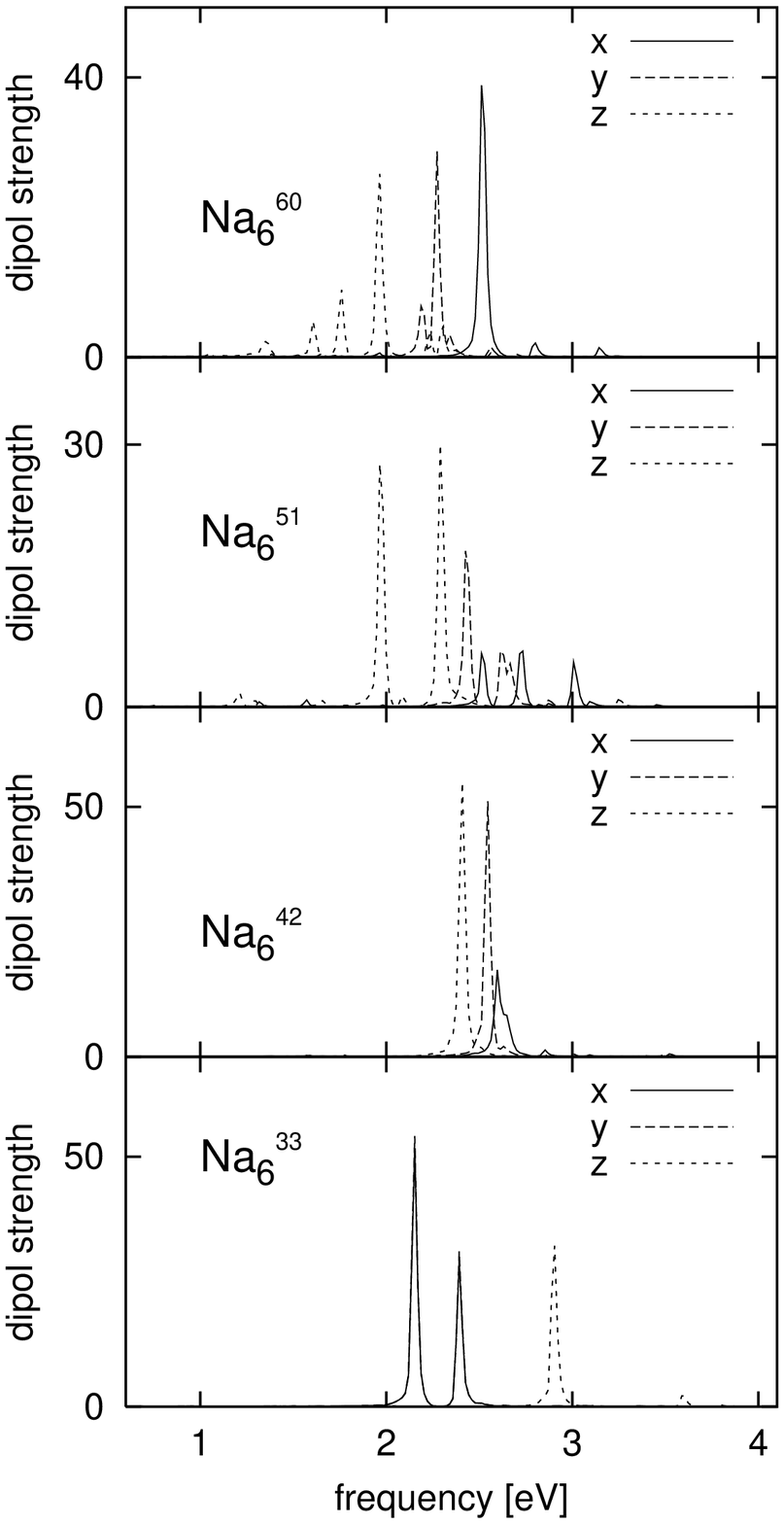,width=7.8cm}} 
\caption{\label{fig:MpNa4} 
Dipole strengths along the three principle axes for the various 
spin-polarized states of Na$_4$ (left) and
Na$_6$ (right) with polarization as indicated by the 
upper index. 
} 
\end{figure}


\subsection{Optical response} 
\label{sec:opt}

 
 
 
It is also interesting to investigate the key feature of cluster
excitations, namely the optical response of polarizd clusters.  Thus
we investigate the dipole strength of the various clusters in our
sample. Thereby we confine considerations to ionic configurations
which are relaxed for given net spin. Figure \ref{fig:MpNa4}
 shows the dipole spectra all in the same manner for a
selection of two clusters and configurations. The dipole strengths are
evaluated along the principal axes which are usually very close to the
optical axes of the clusters (the optical axes are those for which the
dipole response tensor is diagonal). The comparison of spectra can be
helpful to discriminate different spin states where they may
compete in practice. The case becomes of course a bit academic
for the high lying isomers because these are quickly destabilized 
by any additional excitation.

Before starting the detailed discussion, let us briefly recall the 
basic features of dipole spectra in metal clusters 
\cite{ownrw,rpaclust}. The average position is roughly estimated by 
the Mie surface plasmon frequency. The total strength (according to 
the dipole sum rule) is contained to about 90\% in the Mie plasmon 
resonance which can be very well computed in purely collective models 
\cite{rpaclust,locrpa1,locrpa2}.  The remaining strength is found in 
secondary surface plasmons and to a lesser extend in the volume 
plasmon.  There are two mechanisms which produce a spread of the 
spectra around the average surface plasmon frequency.  The global 
quadrupole deformation of the cluster leads to a splitting of the 
plasmon which can still be described at a collective level. The 
oscillations along the elongated axes are red shifted while those 
along the squeezed axes are blue shifted \cite{ekdef}. This 
deformation splitting is an important tool for assessing the 
deformation of small clusters \cite{Sel,Copenhagen,SH}.  The other 
source of spectral broadening is due to the coupling of the resonance 
to nearby one-particle-one-hole $1ph$ state. It is called Landau 
fragmentation because it is the finite systems analogue of Landau 
damping in the electron gas of a plasma. This effect sensitively 
depends on the details of the shell structure. As a rough rule one can 
say that Landau fragmentation becomes increasingly important the 
larger the clusters \cite{rpaclust,sepdef}.  For small clusters it 
plays a role in connection with broken parity symmetry \cite{ownrw}. 
 
In the following we discuss figure \ref{fig:MpNa4}.
We do also comment briefly on the results
and trends obtained for other clusters even if the corresponding
spectra are not displayed here.  There are of course large differences
concerning the degree of splitting or fragmentation, as a function of
size and spin. The cleanest plasmon peaks are found systematically
whenever the magic $N_{\rm spinup}=4$ is involved which happens for
the configurations (not all of them are shown here) ${{\rm
    Na}_{4}}^{40}$, ${{\rm Na}_{5}}^{41}$, ${{\rm Na}_{6}}^{42}$,
${{\rm Na}_{7}}^{43}$, and ${{\rm Na}_{8}}^{44}$. These systems have
all been identified in subsection \ref{sec:config} as being nearly
spherical and accordingly there is no deformation splitting. Moreover,
these configurations seem to be particularly stable as we see no or
very little Landau fragmentation.  The same clean plasmon peak is seen
for the next magic $N_{\rm spinup}=10$ in the spectrum of ${{\rm
 Na}_{10}}^{100}$.
 
More or less fuzzy spectra are seen for the systems without shell 
closures. Let us consider the various pattern of deformation splitting 
and of Landau fragmentation for the various spin states of Na$_6$ 
(and please compare with the shapes as 
shown in figure \ref{fig:na6_detail}).  This example nicely covers all 
variants. The unpolarized cluster ${{\rm Na}_{6}}^{33}$ is axially 
symmetric oblate ($\gamma=60^o$). And accordingly the $x$- and $y$ 
modes are degenerate. They are in the average red shifted relative to 
the $z$ mode. This is the expected deformation splitting.  However, 
one sees two peaks for the $x$-$y$-modes. This is a fragmentation 
caused by coupling to a detailed $1ph$ state.  The ${{\rm 
Na}_{6}}^{42}$ has a magic electron shell. It is nearly spherical and 
correspondingly all modes gather around the same frequency. In fact, 
just for Na$_6$ sphericity is least perfect and thus we see a small 
remainder of deformation splitting.  The cluster ${{\rm Na}_{6}}^{51}$ 
is nearly cylindrically symmetric prolate. Thus $x$- and $y$- modes 
stay close to each other while the center of the $z$-mode is strongly 
red-shifted relative to that. (Remind that it was blue-shifted for the 
oblate case.) All three modes are slightly fragmented by coupling to 
$1ph$ states. Finally the cluster ${{\rm Na}_{6}}^{60}$ is truly 
triaxial ($\gamma=20^0$). And we see indeed a splitting into three 
distinct center frequencies for the three directions. Additionally, 
there is some Landau fragmentation for the $z$-mode. Altogether, we 
see that the simple rules of deformation splitting are well 
observed. The same holds for all other examples. The amount of Landau 
fragmentation is hard to predict in general terms. But the tendency 
that well bound magic shapes have less fragmentation is confirmed. The 
overall trend of average frequency with radius cannot easily be read 
off from the sometimes much split and fragmented spectra. Looking at 
the sequence a bit longer, one can see through the fuzzy pattern the 
expected trend, namely that the largest radius (${{\rm Na}_{6}}^{60}$) 
has lowest frequency and the smallest radius (${{\rm Na}_{6}}^{42}$) 
has the highest frequency. Similar observations can be made for all 
other spectra. 

At second glance, we see in several spectra tiny spots of strength
just above 1 eV. Take, for example, Na$_6$ in figure \ref{fig:MpNa4}.
The spectrum is absolutely empty below 2 eV for the unpolarized state
${{\rm Na}_6}^{33}$. A hint of strength at 1.5 eV shows up in ${{\rm
    Na}_6}^{42}$. It is better visible at the higher spin states where
it finally resides around 1.2-1.3 eV.  Previous studies have shown
that small Na clusters possess spin-dipole modes around 1 eV and that
spin-polarized clusters show cross talks between dipole and
spin-dipole modes \cite{spintdlda,spin3D}.  We have checked that the
small strength above 1 eV seen here is indeed due to cross talk with
the spin-dipole mode. 

Checking from that viewing angle figure
\ref{fig:MpNa4} again, we see that the low lying 
strength in the fully polarized clusters differs very much for the 
various systems in the sample. The point is that the spin-dipole modes 
and the Mie plasmon mode in the dipole channel differ very much in 
their collectivity. Unlike the Mie plasmon mode, the spin-dipole 
experiences only a small residual Coulomb interaction (because the 
shifts of spin-up cloud and spin-down cloud go into opposite 
directions). It resides practically at pure $1ph$ energies. The $1ph$ 
energies, in turn, are determined by the spectral gap at the Fermi 
energy (HOMO-LUMO gap). And this gap depends on shell structure. Magic 
electron numbers, N=4 and 10, have a larger gap than intermediate 
systems. There is no low-lying strength 
in ${{\rm Na}_4}^{40}$ and ${{\rm Na}_{10}}^{10\,0}$ while pronounced 
low lying states appear in the mid shell region.

\section{Conclusions}

We have investigated from a theoretical perspective the properties of 
small Na clusters at systematially varied spin polarization. We used 
as tool density-functional calculations at the level of the 
local-density approximation together with local pseudo-potentials for 
the coupling of Na ions to the valence electrons.  In a first step, we 
have checked possible non-collinearity of the spins. We find always 
fully collinear electron configurations for this simple material Na.

Electronic and ionic structure for small clusters has been discussed 
as well as optical response. Fully spin polarzed clusters display also 
a sequence of magic electron numbers which are just half of the magic 
numbers of spin saturated Na clusters. In our sample, we see the 
impact of the magic $N_{\rm spinup}=4=8/2$ and $N{\rm 
spinup}=10=20/2$.  at various places. Magic $N_{\rm spinup}$ drive the 
system to minimum deformation, minimal radii, and relatively lower 
energies. As a consequence, several clusters show a large 
rearrangement of the ionic configuration when changing spin 
polarization. 
 
We have also investigated optical response because it might provide a 
useful indicator of the underlying spin and ion structure. For the 
small clusters studied here, unpolarized systems show clean Mie 
plasmon resonances with a collective splitting directly related to the 
quadrupole deformation while the spectra of spin-polarized clusters 
show occasionally more fragmentation due to cross talk with spin 
modes. 
 
The question of the life-time for highly spin-polarized clusters 
remains beyond the scope of this paper. It will be attacked in a next 
step.

\bigskip 
 
\noindent 
{\em Acknowledgments: This work has been supported by the French-German 
exchange program PROCOPE, contract number 99074, by  
Institut Universitaire de France,
by the CNRS programe ``Mat{\'e}riaux'' (CPR-ISMIR),
and by a Gay-Lussac prize. The authors furthermore acknowledge 
fruitful discussions with F. Stienkemeier.}

%
%
%
%
%
%


\begin{thebibliography}{99} 
 
\bibitem{toennies} 
J.P. Tonnies, A.F. Vilesov, K.B. Whaley, Phys.Today {\bf 54} (2001) 31 
 
 
\bibitem{doeppner} 
T. D\"oppner, Th. Diederich, J. Tiggesb\"aumker, K.H. Meiwes-Broer",  
Eur.Phys.J. D,  {\bf 16} (2001) 13\\ 
 
 
 
 
\bibitem{leiderer} 
P. Leiderer,  Z.Phys. B {\bf 98} (1995) 303\\ 
 
\bibitem{anciletto} 
F. Anciletto, E. Cheng, M.W. Cole, F. Taigo. Z.Phys. B {\bf 98} (1995) 323\\ 
 
\bibitem{Sti01} 
F. Stienkemeier, A.F. Vilesov, J.Chem.Phys. {\bf 115} (2001) 10119 
 
 
\bibitem{stienke} 
C. P. Schulz, P. Claas, D. Schumacher, and F. Stienkemeier,   
Phys.Rev.Lett. {\bf 92} (2004) 013401 
 
 
 
\bibitem{magnresp} 
C. Kohl, B. Fischer, P.-G. Reinhard, Phys.Rev. B {\bf 56} (1997) 11149 
 
\bibitem{magnopt} 
C. Kohl, S.M. El-Gammal, F. Calvayrac,  E. Suraud, P.-G. Reinhard, 
Eur.Phys.Journ. D {\bf 5} (1999) 271 
 
 
\bibitem{perwan} 
J.\ P.\ Perdew and Y.\ Wang, Phys.\ Rev.\ B \textbf{45}  (1992) 13244 
 
 
 
\bibitem{adsic} 
C. Legrand, E. Suraud, P.-G. Reinhard,  
J. Phys. B {\bf 35} (2002) 1115 
 
 
\bibitem{kuemmel}  
S. K\"ummel, M. Brack, P.-G. Reinhard,  
Eur. Phys. J. D {\bf 9} (1999) 149  
 
 
\bibitem{Koh99b} 
C. Kohl, G.F. Bertsch, Phys.Rev. B {\bf 60} (1999) 4205 
 
 
\bibitem{lauritsch} 
V. Blum, G. Lauritsch, J.A. Maruhn, P.-G. Reinhard, 
J. Comp. Phys. {\bf 100} (1992) 364 
 
 
\bibitem{Metro}  
C. Kohl, PhD thesis, Erlangen 1997 
 
\bibitem{Fei82} 
M.D. Feit, J.A. Fleck, A. Steiger, J.Comp.Phys. {\bf 47} (1982) 412 
 
 
\bibitem{yabber} 
K. Yabana, G.F. Bertsch,  Z.Phys. D {\bf 42} (1997) 219 
 
 
\bibitem{bigtdlda} 
F. Calvayrac, E. Suraud, P.-G. Reinhard,  
Ann.Phys. {\bf 254} (N.Y.) (1997) 125 
 
 
\bibitem{ownrw} 
F.  Calvayrac, P.-G.  Reinhard, E. Suraud, C. Ullrich, 
Phys.Rep. {\bf 337} (2000) 493 
 
 
 
\bibitem{mbrev} 
M.\ Brack, Rev.Mod.Phys. {\bf 65} (1993) 677 
 
\bibitem{CAPS} 
B. Montag, P.-G. Reinhard, Phys.Rev. {\bf B51} (1995) 14686 
 
\bibitem{rpaclust} 
P.-G. Reinhard, O. Genzken, M. Brack, 
Ann.Phys. (Leipzig) {\bf 5} (1996) 576 
 
\bibitem{sajm} 
B. Montag, P.-G. Reinhard, J. Meyer, 
Z.Phys. {\bf D32} (1994) 125 
 
\bibitem{caps55} 
S. K\"ummel, M.Brack, and P.-G. Reinhard, 
Phys.Rev. B {\bf 62} (2000) 7602 
 
\bibitem{mgexp} 
T. Diederich, T. D\"oppner, J. Tiggesb\"aumker, K.-H. Meiwes-Broer, 
Phys.Rev.Lett.\ {\bf 86} (2001)  4807 
 
 
\bibitem{sapsmg} 
Ll. Serra, P.-G. Reinhard, E. Suraud,  
Euro.Phys.J. D {\bf 18} (2002) 327 
 
\bibitem{locrpa1} 
M.\ Brack, Phys.\ Rev.\ B \textbf{ 39} (1989) 3533 
 
\bibitem{locrpa2} 
P.-G. Reinhard, M. Brack, 
Phys. Rev. {\bf A41} (1990) 5568 
 
\bibitem{ekdef} 
W. Ekardt, Z. Penzar, Phys.Rev. B {\bf 43} (1991) 1331 
 
\bibitem{Sel} 
   K. Selby, M. Vollmer, J. Masui, V. Kresin, W.A. de Heer and 
   W.D. Knight, Phys. Rev. B{\bf 40} (1989) 5417 
\bibitem{Copenhagen} 
    P. Meibom, M. $\oslash$sterg${\ddot \rm a}$rd, 
    J. Borggreen, S. Bjornholm and  
    H.D. Rasmussen, Z. Phys. D{\bf 40} (1997) 258 
\bibitem{SH}    
   H. Haberland and M. Schmidt,  
   Eur. Phys. J. D{\bf 6} (1999) 109 
 
\bibitem{sepdef} 
V. O. Nesterenko, W. Kleinig, P.--G. Reinhard, 
Euro.Phys.J. D {\bf 19} (2002) 57 
 
\bibitem{spintdlda} 
L. Mornas, F. Calvayrac, P.-G. Reinhard, E. Suraud, 
Z.Phys. {\bf D38} (1996) 73 
 
\bibitem{spin3D} 
C. Kohl, S.M. El-Gammal, F. Calvayrac, E. Suraud, P.-G. Reinhard, 
Eur.Phys.Journ. D {\bf 5} (1999) 271 
 
 
\bibitem{spinjel} 
C. Kohl, B. Montag, P.-G. Reinhard, Z. Phys. D {\bf 35} (1995) 57 
 
\bibitem{Gar91} 
M.E. Garcia, G.M. Pastor, K.H. Benneman, 
Phys.Rev.Lett. {\bf 67} (1991) 1142 
 
 
\end{thebibliography}
\end{document}